\documentclass[reprint, amsmath,amssymb, aps,]{revtex4-2}
\usepackage[english]{babel}
\usepackage{graphicx}
\usepackage{dcolumn}
\usepackage{bm}
\usepackage{braket}
\usepackage{import}

\usepackage{dsfont}
\usepackage{amsfonts}
\usepackage{dirtytalk} 

\usepackage{tikz}
\usetikzlibrary{positioning}
\usetikzlibrary{shapes,snakes}
\usetikzlibrary{graphs, graphs.standard}

\usepackage{tabularx}
\usepackage[math]{cellspace}

\usepackage{pifont}

\usepackage{hyperref}
\usepackage{url}

\newcommand{\xmark}{\text{\ding{55}}}%
\newtheorem{theorem}{Définition}
\newtheorem{proposition}{Proposition}
\newtheorem{definition}{Definition}

\providecommand{\moy}[1]{\langle #1 \rangle}
\newcommand{\pmat}[4]{\begin{pmatrix} #1 & #2 \\ #3 & #4\end{pmatrix}}
\DeclareMathOperator{\Tr}{Tr}

\begin{document}

\title{From the problem of Future Contingents to Peres-Mermin square experiments:\\ An introductory review to Contextuality}

\author{Ga\"el Mass\'e}
 \altaffiliation[]{Laboratoire Mat\'eriaux et Ph\'enom\`enes Quantiques.}

\affiliation{%
 Laboratoire Mat\'eriaux et Ph\'enom\`enes Quantiques, Universit\'e de Paris}%

\date{\today}

\begin{abstract}
A review is made of the field of contextuality in quantum mechanics. We study the historical emergence of the concept from philosophical and logical issues. We present and compare the main theoretical frameworks that have been derived. Finally, we focus on the complex task of establishing experimental tests of contextuality. Throughout this work, we try to show that the conceptualisation of contextuality has progressed through different complementary perspectives, before summoning them together to analyse the signification of contextuality experiments. Doing so, we argue that contextuality emerged as a discrete logical problem and developed into a quantifiable quantum resource.
\end{abstract}

\maketitle

\tableofcontents

\section*{Outline}
The fact that in classical physics the objects that possess physical properties and the objects that are used to formulate predictions regarding the evolution of these physical properties are confounded comforts the belief that this distinction does not exist~\cite{Bitbol2015, Bitbol2008, Bitbol2014}. Yet such a distinction is mathematically translated in the principles of Quantum Theory - the evolution of an observable being mediated by a prospective calculation of probability~\cite{PeresZurek1982, Laloe2001, Peres84}. Hence, a fundamental question regarding quantum mechanics is the possibility to assign preexisting values to an object before a measurement is done, in opposition to the assumption that the value is brought into being by the probing apparatus. If this was possible, one could determine the measurement result of an observable independently of the other observables measured during the measurement setup (the context of measurement). The notion of contextuality refers to the impossibility of such an assignment. Its study began with philosophical considerations and reached a first milestone when it was mathematically proven that quantum mechanics was contextual. This will be the subject of the first part of this chapter, called \textit{An historical and philosophical introduction to the origins of contextuality} made of sections~\ref{OriginsContextuality} and \ref{MathematicalProofs}. Then, structural theories were derived, whose purposes were to generalise the previous works, sometimes beyond quantum mechanics, and deepen the conceptual understanding of the notion. This will be treated in a second part called \textit{The structural theories} which encompasses sections~\ref{CSW}, ~\ref{Abramsky},~\ref{ALFS}~\ref{SpekkensContextuality},~\ref{Cbd} and~\ref{comparisons}. Finally, in the third part, \textit{The quest to derive experimentally robust proofs of contextuality}, we will study, in the light of the various approaches of the structural theories, the conceptual problems that arise when one wishes to adapt these proofs to realistic set up (section~\ref{Analysis}), before presenting and deriving a road map to tackle them, in sections~\ref{NewBounds},~\ref{NoiseforPM} and~\ref{FinalProtocol}. We start by a general introduction to contextuality and correlation experiments.

\section{Introduction}
\label{Introduction}

\subsection{Measurement is an invasive process}
\label{sub:Measurement}
 One of the most iconic experiments in the history of quantum physics is the double-slit experiment. Its most fundamental version is displayed by the following protocol: one uses a coherent source of light, such as a laser to illuminate a plate pierced by two parallel slits and observes the light passing through the slits on a screen. An interference patterns appears, characteristic of the wave aspect of light. However, if photon detectors are placed at the slits location, photons are recorded to go through either one or the other split, and not both at the same time: this is characteristic of the particle aspect of light. Besides, the light pattern is changed so that the interference pattern vanishes. If we modulate the efficiency of the detectors, the photons that are detected will contribute to a particle-like pattern, and the undetected ones to a wave-like pattern.
This experiment has been replicated with electrons, atoms and molecules and is a clear demonstration of the wave-particle duality of light and matter. From a quantum information perspective, we learn the following lesson: no detector has ever been able to be at the same time sensitive enough to detect a photon, and sufficiently smooth to avoid disturbing it. In other terms, any acquisition of information leads to a disturbance of the system. This technological impossibility has been turned into a fundamental principle of quantum mechanics, according to Feynman~\cite{Feynman2006}: the act of measurement is an invasive process which modifies the system. This is mathematically displayed by the Born rule which yields the probabilities of obtaining an outcome and the Lüder's rule that gives the modification of the system after a measurement, given a complete knowledge of it. 

\subsection{Beyond measurement disturbance}
\label{sub:Disturbance}
Is this perturbation enough to ensure that measurements outcomes can not be assigned predetermined values? No, we could imagine that observables have a deterministic response to measurements that are revealed by this process, even if the system is disturbed afterwards~\cite{Mermin1993}. But then, wouldn't it be possible to invoke the uncertainty principle? Since there exist observables that can not be jointly measured, how could there be predetermined values that induce the answer to a measurement, without taking the disturbance of other observables into account? In fact, the uncertainty principle prevents us from assigning infinitely sharp values to a set of observables, but does not say anything for an observable alone~\cite{Mermin1993}. To say it differently, the fact that we can not measure simultaneously the position and the velocity of an electron does not mean that they do not exist independently one from the other. If we want to prove this, we need to make these hypothesis more explicit, and that is the very purpose of hidden variable models. We mention that this approach, favoured by Schrödinger, de Broglie and Einstein and deemed sceptically by Heisenberg, Born, Pauli and Bohr, has been a source of tremendous controversy, and refer the reader to~\cite{Bohr1958, Cabello2020conf, Bacciagaluppi2009, Becker2015} for further information. 

\par In order to isolate the effect of predetermined assignment, we require to measure an observable only with other observables that do not disturb it. These are called compatible observables, and they can be jointly, simultaneously, measured. When we measure sequentially a set of compatible observables, the order of measurement has no impact on the results. Let us now measure one of these observables, but within another measurement context, \textit{i.e.}, with another set of compatible observables. The fact that, in some cases, this observable takes different values than in the previous context proves that quantum mechanics is contextual. One can not predict the outcome of a measurement without having taken into account, not only the original state of the system, but also all the other measurements. It is not possible to define an observable without reference to the context inside which it is measured.

\subsection{General considerations on correlation experiments}
\subsubsection{Definition of an abstract experiment}
\label{subsub:CorExp}
In order to experimentally investigate contextuality, \textit{correlation experiments} have been conceptualised. They are defined by scenarii in which different observers can perform measurements on systems and record outcomes, before they compare them and compute their correlations. We assume that the different experimental protocols, in which the systems can be made of ions, photons or atoms, and the measurement being led with photodetectors, homodyne detectors or others are just particular realisations of a unique abstract experiment. The physical systems are assumed to be reproducible~\cite{CSW2014, Spekkens2005}, and the only relevant data are thus the statistics extracted from the experiments. The outcomes of the measurements are distributed according to joint probability distribution, denoting the probability that observers obtain results given they have performed some measurements. 

\subsubsection{Illustration}
\label{subsub:illustrationcorr}

 \par
\begin{figure}[h]
  \centering
  \def\svgwidth{\linewidth}
  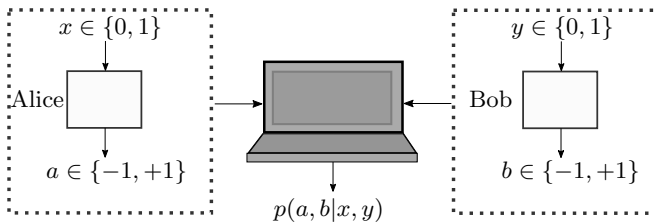
  \caption{A bipartite correlation experiment}
  \label{fig:Bell}
\end{figure}

 We illustrate this general concept with a particularly simple scenario. In this particular setup, where we use the Bell scenario terminology, two observers, Alice and Bob, can perform measurements on different systems. Alice can choose between two measurements $x \in \{ 0,1 \}$, and for each measurement she obtains two possible outcomes that are denoted by $a \in \{-1,1\}$. Similarly, Bob can choose two measurements $y \in \{ 0,1 \}$ with possible outcomes $b \in \{-1,1\}$. The choices and results are then compared. The joint probability distribution computed out of it will be denoted $p(a,b|x,y)$. This scenario is depicted in Figure~\ref{fig:Bell}. The general scenarii that we have described previously are just generalisation of these kind of setups, where the number of different observers $o$, measurements $m$ and possible outcomes $d$ can take any discrete finite value.

\subsubsection{Protocols to explain the correlation experiments}
\label{subsub:correlationsprotocols}
We try to understand better these correlation experiments by confronting them to different models. It is indeed necessary, not only to propose a model that is able to reproduce the experimental results, but also to exhibit models that can not, in order to understand which axioms of the models are necessary and which ones are not. That is why hidden-variable programs have been established. They are made of models derived from minimal and intuitive hypothesis. Their adequacy with empirical data results provides information on the validity of their hypothesis. The hypothesis are in fact evaluated, not only on their capacity of prediction, but also on their capacity to deepen our understanding and to expand our theories.

\subsubsection{From correlations to causal explanations}
\label{subsub:correlationsexplanations}
The quantum mechanics framework does not contain spatial or temporal limitations to these correlations. However, the predictive capacity of this theory does not satisfy all physicists. There is a general belief that understanding these correlations implies giving them a causal explanation, and to do so, if we reject the vague and limited concepts of harmony and pure coincidence, we are seemingly left with a choice between a mutual influence from a distance, and the existence of a preexisting common cause. Yet, the common underlying hypothesis in both cases is the idea of isolated experimental \textit{events}, which would be triggered and updated. This notion may lead to logical dead-ends, and could be rejected and replaced by a notion of global coherence where no individual objects may bear signification~\cite{Bitbol2015}.

\subsubsection{No-signalling principle}
\label{subsub:NS}
The no-signalling principle states that the outcomes obtained by an observer is independent from the choice of measurement made by another observer. In Bell-scenario, a kind of correlation experiment where observers are space-like separated~\cite{Brunner2014}, it denies the possibility that different elements of the experiment communicate above light speed. Mathematically, in the paradigmatic experiment we presented in Figure~\ref{fig:Bell}, the no-signalling constraints are expressed as constraints on the marginals of Alice and Bob:
\begin{equation}
    \sum_{b=1}^{m-1} p(a,b|x,y) = \sum_{b=1}^{m-1} p(a,b|x,y'), \forall \ a,x,y,y'
    \label{eq:NSAlice}
\end{equation}
\begin{equation}
    \sum_{a=1}^{m-1} p(a,b|x,y) = \sum_{a=1}^{m-1} p(a,b|x',y), \forall \ b,x,x',y.
    \label{eq:NSBob}
\end{equation}
We provide a clarification of the two levels of this concept due to Sainz~\cite{SainzConv}:
\begin{enumerate}
    \item There is a probabilistic model $p(a,b|x,y)$ characteristic of the device shared by Alice and Bob through its input/output functionality and the statistical data that they obtain.
    \item There are the tasks undertaken by Alice and Bob  when they share such a device.
\end{enumerate}
The device must obey non-signalling conditions when the tasks that Alice and Bob are allowed to do are communication scenarii. Let us however mention that a relaxation of this condition has been studied within the framework of Bell scenarii~\cite{Brask2017}. In fact, we can (we will explore it in this work) have a quantitative approach to the no-signalling principle~\cite{Hall2011}.
\par The correlations that are not excluded by the no-signalling principle form a polytope in the whole space of all possible correlations, which becomes more and more difficult to characterise as the number of observers, measurements and outcomes grows~\cite{Pironio2011}. The no-signalling condition is very general and has been adapted to other fundamental tests of quantum mechanics~\cite{Leggett1985}, turning from a spatial to a time condition~\cite{Kofler2013, Guryanova2019}. Some analysis of contextuality scenarii presented in this work use this principle~\cite{Abramsky2017, Acin2015, Kujala2015}, in a manner that we will discuss. It will play a central part as far as the experimental possible loopholes are concerned. It has been shown to be a resource for cryptography~\cite{Barrett2005} and private randomness~\cite{Colbeck2009} when added to contextuality.

\subsection{Relation to nonlocality}
\label{sub:relationlocality}
The field of contextuality has deep links with the field of Bell nonlocality~\cite{Brunner2014}, the fact that the predictions of quantum mechanics can not be accounted for by any local theory. These similarities were perceived very early~\cite{Mermin1993} and are still investigated today~\cite{Suarez2017, Cabello2019, Spekkens2014}, some recent works perceiving contextuality as a generalisation of nonlocality~\cite{Silva2017}. Unlike nonlocality, contextuality does not need two separated sites to manifest its effects. Both fields focus on correlation experiments and the possibility to emulate the statistics obtained with well chosen hidden variables. In nonlocality scenarii, the crucial assumption of the hidden variable models) is one of factorisability, whereas in noncontextual ones, it is one of determinism. Several mathematical constructions encompasses directly encompasses both fields~\cite{CSW2014}. In particular, under the sheaf approach, it has been shown that in any experimental scenario, if a factorisable hidden variable model holds, then a noncontextual one does also, and reciprocally~\cite{Abramsky2011}. This result has been extended to continuous variables and has been called the Fine-Abramsky-Brandenburger theorem~\cite{Barbosa2019}, as it built on a previous similar result from Fine, restricted to the CHSH scenario~\cite{Fine1982}.

\subsection{Detailed plan}
\label{sub:Outline}
In section~\ref{OriginsContextuality}, we give a philosophical and historical introduction to the concept of contextuality in quantum physics. We present the philosophical and scientific context in which the concept emerged, and deliver an analysis on its links with the structure of the contigent futures problems. We shortly present the mathematical proofs of quantum contextuality, from the Kochen-Specker theorem~\cite{KochenSpecker1967} to its extension to non-contextuality inequalities like the so-called KCBS inequality~\cite{Klyachko2008} and the Peres-Mermin Square~\cite{ Mermin1993} in section~\ref{MathematicalProofs}. Once this is done, we turn to a systematic presentation of what we call the structural theories of contextuality. We begin in section~\ref{CSW} by presenting the graph theoretic approach that generalises the previous proofs~\cite{CSW2014}, and then in section~\ref{Abramsky} a sheaf-theoretic approach that unifies the study of contextuality and non-locality and hierarchies contextuality~\cite{Abramsky2011}. We then discuss in section~\ref{ALFS} to what extent hypergraph approach of~\cite{Acin2015} subsumes them both. However, to understand the relevance of this association, we will present an operational theory of contextuality~\cite{Spekkens2005} in section~\ref{SpekkensContextuality}. It noticeably defines contextuality beyond quantum mechanics and highlights the conceptual problems that arise when seemingly experimentally robust noncontextuality inequalities are derived inside quantum theories. The last conceptual approach to correlation experiments, named Contextuality-by-Defaut (CbD)~\cite{Dzhafarov2015}, will give us a road map to control the experimental loopholes. We discuss a comparison of all these approaches in section~\ref{comparisons}. We then move to the final part, in which we summon all these works to properly explain the meaning of experiments of contextuality. We begin by explaining the different loopholes~\cite{Meyer1999, Spekkens2014} that have been highlighted in the field of Quantum Contextuality, and discuss their scope in~\ref{Analysis} We then explicitly treat one particular setup and check that one of its experimental realisation indeed witnessed contextuality: the Peres-Mermin square. We review the different ways to study its contextuality from the graph theoretic and operational framework, before reinterpreting the results of the Kirchmair experiment~\cite{kirchmair2009} in the light of the sheaf-category and the CbD approach in section~\ref{NewBounds}. We argue that even if these results are sufficient to highlight the presence of contextuality, it is nonetheless relevant to derive a purely QM model that reproduces the result of experiments. We present it in section~\ref{NoiseforPM}, notably reviewing and discussing the models of Szangolies~\cite{Szangolies2015}. We finally conclude this part in section~\ref{FinalProtocol} by submitting a protocol that encompasses the precedent results. Finally, we conclude our work in~\ref{FinalProtocol} with some openings on the use of contextuality. We took great support from the presentation of Hippolyte Dourdent~\cite{Dourdent2018} (for the French readers only) throughout this thesis.

\subsection{The purpose of this article - disclaimer}
\label{sub:disclaim}
The initial goal of this work was to present models that could explain the experimental data in contextuality experiments. In order to do so, it soon appeared necessary to the author to clarify and discuss the links and scope of the multiple points of view on contextuality that had been developed. Recognizing four (five with the hypergraph) major approaches to contextuality, I chose to focus on one particular experiment, the Peres-Mermin square, as a paradigmatic example to introduce the multiple answers to the controversies about the possible loopholes that had been discussed in the last decades. It seemed fruitful to review in detail, in a previous part, these approaches for three reasons:
\begin{enumerate}
    \item Presenting a self-contained review
    \item Clarifying the links between the different theories
    \item Producing an introductory review to the field
\end{enumerate}
Indeed, at the time when this work was undertaken, no such review existed. I sought to avoid a long-winded effort to colleagues that would begin in the field by displaying my own understanding in a pedagogical manner. This may explain the large number of sources such as video conferences quoted here (notably from the Perimeter Insitute Video Library, or the QCQMB Colloquium youtube page), as they were for me a much-needed introductory step towards the subtleties of contextuality. Also, it justified an interesting comparison between the different approaches. 
\par Few time before publishing this work, a review called \textit{Quantum Contextuality} was uploaded online~\cite{Budroni2021}. After some discussion with one of its authors, I chose to keep the review form, even if it intersects parts of their work, since the experimentally- and significance- oriented approaches seemed to offer a different understanding. Indeed, the will to catch intuition and to gain understanding on the subject had also led to extend this work to a careful analysis of the emergence of the concept of contextuality in quantum mechanics, from an historical and a philosophical perspective, as well as from the logical structure of the problems raised by that notion. With a few last words on the future of contextuality and the possibility it offered, notably from a computational aspect, this mini-thesis was complete.  
\par It is intended to be a brick in the wall of the collaborative project that is the study of contextuality. I tried my best to encompass many different studies that might help the reader, whether they come from the fields of logic, philosophy, history, quantum mechanics or mathematics. I would gladly receive any comment, remark, criticism or advice on this work, which may in turn be integrated.

\clearpage
\begin{center}
\textbf{\textsc{\Large An historical and philosophical introduction to contextuality}}
\end{center}


\section{Origins of contextuality}
\label{OriginsContextuality}

\subsection{The influence of Gonseth}
\label{sub:Philo}
Specker published in 1960  an article entitled \textit{The logic of non-simultaneously decidable propositions} \cite{Specker1960}, which preluded to his famous result, in \textit{Dialectica}, a journal about scientific and cognition philosophy founded by Gaston Bachelard, Paul Bernays and Ferdinand Gonseth. Gonseth had been the teacher of Specker and an inspirational figure for him~\cite{Dourdent2018}. He developed the idea that logic was based on the existence of objects on which propositions apply, against the neo-positivism of the then influential Vienna Circle~\cite{Emery1985} (see section~\ref{sub:neopositivism} below). Indeed, from his own reflection upon his study of philosophy, mathematical foundations, physics as well as biology, he gained the perception that any attempt to establish a formal, logical meta-theory that would constitute an irrefutable foundation of science  would fail~\cite{Tannoudji2004}, a position reinforced by the then recent discovery of Gödel's incompleteness theorem~\cite{Godel1931}. Not only would this attempt dangerously hide the intuition and the experiments from which axioms are deduced, and consequently reduce their scope, but axioms and logic are not so different that it is justified and fruitful to separate them~\cite{Emery1985}. Rather, science dialectically progresses by a constant dialogue between them. This will be notably illustrated with the example of the mathematical foundations by Imre Lakatos~\cite{Lakatos1976}. Gonseth summed up his views in the provocative declaration \textit{Logic is in the first place a natural science}~\cite{Gonseth1936} that Specker quoted in the epigraph of his article.

\subsection{A metaphysical motivation}
\label{sub:counterfactuals}
Ironically, Gonseth deduced from his concept of logical objects the three pillars of classical logical, the law of excluded middle, the law of Non-Contradiction and the law of identity, in the form of the three following conceptions of objects of logic~\cite{Dourdent2018}:
\begin{itemize}
    \item Any object is or is not
    \item An object cannot be and not be at the same time
    \item Any object is identical to itself,
\end{itemize}
propositions shown to be incompatible with the rules of quantum logic.
\par Specker was interested in the logical problems that could emerge when one wished to attribute a truth value to a proposition about the future, such as the famous example introduced by Aristotle: \textit{A sea battle will take place tommorow}. In order to avoid paradoxes, Aristotle stated that propositions about the future were \textit{contingent}, that they could only be attributed a truth value when they become actual~\cite{Dourdent2018}. The study of \textit{Future Contingents} has long been linked with the study of \textit{counterfactual propositions}, propositions of conditional logic based on unverified hypothesis~\cite{Dourdent2018}. Their relations to contextuality are still studied nowadays~\cite{Svozil2005}. In the XIII century, Thomas Aquinas raised these questions about counterfactuality into theological questions. In \textit{Summa Theologica}, he analysed the questions: \textit{Does God have the knowledge of things that are not? Does God know the contingents futures?}. Interestingly, if Thomas and his adversaries were mostly concerned with the metaphysical aspect of these problems, seeking to reconcile the omniscience of God with human's free will~\cite{Anfray2004}, we also find interesting from a scientific point of view to analyse the different logical solutions that were proposed throughout the centuries, in order to highlight the specific aspects of the quantum framework and the place of contextuality inside it. Indeed, with the assumption that “something is true if and only if it is known to God", a formal connection is established between the theological questions and the logical ones~\cite{FutCon2020}. Specker had recognized this link, and translated it in terms of quantum mechanics by wondering whether it would be possible to know the result of an unperformed measurement, without having a contradiction (according to~\cite{Dourdent2018}, based on~\cite{Spekkens2011}). (Note that metaphysical motivations are still assumed explicitly today, as Kochen published along with Conway two versions of a theorem closed to the Kochen-Specker original theorem, called “The Free Will theorem" and the “Strong Free Will theorem"~\cite{Conway2006, Conway2009}.)

\subsection{Description of the problem of future contingents}

\par Before going back to Specker work in itself, we would consequently like to make a short review of the historical response to this old problem, and to classify them according to the logical rules they rest on. After that, we will evaluate the conformity of these non-classical logical frameworks to the case of contextuality in the framework of quantum mechanics. This section is mostly a sum-up of 1.C and .D from the review of \textit{The Internet Encyclopedia of Philosophia}~\cite{FutCon}. Let us mention that the French reader may consult~\cite{Anfray2004} for an historical review of the philosophical problems implied by the problem of future contingents.

\par Consider the proposition 
\begin{itemize}
    \item (A) \emph{There will be a sea battle tomorrow}
\end{itemize} and its negation
\begin{itemize}
    \item ($\bar{A}$) \emph{There will not be a sea battle tomorrow}.
\end{itemize}
We are tempted to adopt the \textit{Bivalence principle} which states that a proposition is either true, or false. A fatalist, who links truth to time outside of causal relations~\cite{Anfray2004} infers necessity from truthness and impossibility from falseness (principle called (F)). For her, everything is either necessary of impossible. In that case, if (A) is true and ($\bar{A}$) is false, then the sea battle is necessary, if (A) is false and ($\bar{A}$) is true, then the sea battle is impossible. However, this event (the sea battle) is contingent. Consequently, if we refuse to grant a truth value to our sentences at this stage, we need to consider that neither (A) nor ($\bar{A}$) are true or false. They are indeterminate. We apply the principle of Excluded Middle (EM) which states that the disjunction of a proposition and its negation is always true.  Hence, we say that:
\begin{itemize}
    \item \emph{There will or there will not be a sea battle tomorrow}
\end{itemize} is true.  

\par So far, have presented two logical and one metaphysical principles~\cite{FutCon}. Let us write them logically:
\begin{itemize}
    \item (B) Either 'p' is true or 'p' is false
    \item (E) Either 'p' or 'not p'
    \item (F) Either it is necessary that 'p' or it is impossible that 'p'
\end{itemize}
where 'p' stands for any proposition (in our example, 'p' is (A)). The problem of Future Contingent stems from the fact that in classical logics, Excluded Middle implies Bivalence and that, with the adjunction of the hypothesis “if 'p' is true then 'p' is necessary", Bivalence implies Fatalism. Consequently, Excluded Middle would entail Fatalism. Rejecting Fatalism would imply to reject Excluded Middle. Most philosophers would refuse this alternative~\cite{Anfray2004, Jacotte}. In order to answer this problem, they put forward that propositions about the future should be treated inside non-classical logics. Specker wanted to solve this problem, and analysed it inside the quantum framework. Hence, the problem of contextuality can be seen loosely as the problem of future contingents inside the quantum framework. 

\par In the same time that the field of non-classical logics  boomed, thanks notably to the use of modern logic symbolism~\cite{FutCon2020}, Quantum Mechanics was explicitly analysed inside a non-classical and new Quantum Logic (QL)~\cite{Birkhoff1936}. We introduce non-classical logics thanks to the problem of future contingents in the next subsection, before reviewing, in their light, the logical analysis of contextuality inside quantum logic in subsection~\ref{quantumlogic}. Let us mention that links between future contigent and quantum mechanics experiments have been directly assessed, thanks to the quantification of the degree of truth of a proposition based on the probability of its occurence~\cite{Sudbery2017}.

\subsection{Classification of the possible answers}
We shortly review the different historical propositions,  classified according to the four possible logical positions on the problem: the acceptance or not of the Excluded Middle and Bivalence (for a clarification on the difference between them, see~\cite{Beziau2003}). Just like~\cite{FutCon}, we however do not mention Bivalence without Excluded Middle, this option being neither interesting philosophically neither relevant in the framework of the quantum framework. 

\subsubsection{Excluded Middle and Bivalence}
The first solution distinguishes between a truth and a necessary truth. This enables to assume Excluded Middle and Bivalence, while at the same time not accepting Fatalism since this last one is not implied by Bivalence. It was initially proposed by William of Ockham in the XIV century before being formalised and developed notably by Prior, who proposed a branching times semantics (different futures exist) and a truth function that is itself a function of time. 
His logical model is a partially ordered set of moments of time, where truth can be attributed to a a set of histories, called chronicle (a maximally ordered linear subset of possible histories)~\cite{FutCon2020}. A lot of variations on these perspectives have been proposed, notably some that discuss the implications of a true future, a privileged line of time, which is clearly the weak point of this solution~\cite{FutCon}. 

\subsubsection{Excluded Middle without Bivalence}
The rejection of the notion of true future, motivated by philosophical or logical reasons, has led to the development of the so called Supervaluatioanist theory due to Thomason. Interestingly, in this version, a proposition is true at a time t if and only if it is true in all chronicles passing through t. Future contingents do not meet this requirement, so they are indeterminate. In this version, we obtain one of the main requirement of quantum logic (see below), namely that a disjuncted proposition can be true whereas neither of its subparts are~\cite{FutCon2020}. I formulate the hypothesis that a formal link can be established between the logical concept of chronicles in these tense-logical models, and that of contexts in the quantum framework. MacFarlane has suggested another approach of solving the problem of future contingents, by asserting that a proposition can be deemed truth according to both the context in which the proposition is made and that in which it is received. This Relativist version has in common with the Supervaluationist that they both accept the Excluded Middle and reject Bivalence. Let us finally note that Supervaluationism has been explicitly studied as a possible semantic for quantum logic~\cite{Bolotin2017Super}(see also few lines below).

\subsubsection{Neither Bivalence nor Excluded Middle}

Another interesting option for our perspective has been developed by one of the most influential logician of the XX century, Jan Lukasiewicz. He gave an interpretation of the future contingent problem by proposing a many-valued logic, one where propositions can take three values: truth, falsity and indeterminacy~\cite{FutCon}. It however shares with classical logic the tenet that the value of a complex proposition is determined by the values of its constituents. This newly formalised logic was to see its importance grow more and more, especially through its development by Lukasiewicz and Tarski, and later by Hajek inside an infinite-valued, or fuzzy logic~\cite{Pelletier2000}, the utility of which we mention in the next section. 

\subsection{Contextuality in terms of Quantum Logic}
\label{quantumlogic}
\subsubsection{What is Quantum Logic?}
Quantum Logic(s) is a propositional structure constructed in order to  describe appropriately the relations between the events of interest within the quantum formalism~\cite{QuantLogicIEP}. Indeed, the Boolean algebra could not adequately assign truth values to experimental propositions about the position and momentum of a system and to their disjonction. Strictly speaking, quantum logics are generally orthomodular partially ordered set with an ordering set of probability measures~\cite{Pykacz2010, Dalla2002}. We may intuitively approach it by considering that it is a weaker structure than classical logic, for which the most notable aspect is the relaxation of the distributive properties of conjunction and disjunction~\cite{QuantLogicIEP}. We shall not enter here further in the technical details of QL. The reader may consult~\cite{Coecke2000} as an introduction, and~\cite{Dalla2002} for a more comprehensive review. 

\subsubsection{The logic of contextuality}
It has been proven recently that the Kochen-Specker theorem implies a logic that does not contain the principle of bivalence~\cite{Bolotin2018Bivalence, Bolotin2017relation, Bolotin2018truth}, in the sense that it is not possible to provide a truth value to all propositions without contradiction, unless a many-valued logic is chosen. This is consistent with the contextuality of quantum mechanics (for which quantum logic does not respect bivalence~\cite{Dalla2002}) and also restricts the class of possible hidden variable models~\cite{Bolotin2018Bivalence}. More subtle is the analysis of the Excluded Middle. In~\cite{Bolotin2018EM}, it is shown that it can fail for a quantum proposition concerning a qubit, while its space is isomorphic to the lattice of all the closed subspaces of a Hilbert space where it holds. Possible solutions are then envisaged, leading to a supervaluationist, an intuistic or a many-valued logical approaches to quantum logic. 

\subsubsection{A fruitful logical analysis} 
Interestingly, it was shown that every quantum logic can be treated as a partial infinite-valued Lukasiewicz logic. This unifies two distinct understanding of QL, one in terms of many-valued logic, and the other in terms of two-valued but non-distributive~\cite{Pykacz2010, Pykacz2019} that we mentioned above. Accordingly, we join J. Pykacz  and Sudbery on arguing that both the propositions about future contingents and those about uncertain quantum predictions can be adequately treated in the same framework, that of many valued logics~\cite{Pykacz2015, Pykacz2010, Sudbery2017}.  Indeed, if some initial ambitions of the Quantum Logic research programs (namely, to solve all quantum paradoxes) have been abandoned or revised~\cite{Bacciagaluppi2009logic}, it is nonetheless now believed that adapting the structure of logic according to the experimental context studied is legitimate~\cite{Dalla2002} and possibly fruitful. According to Dalla Chiara and Giuntini, the relevant question is now more about choosing the right logic in every context, rather than arguing in a favor of a unique one. In that perspective, Abramsky and Barbosa have recently build a strong connection between partial Boolean algebras (which slightly differs from traditional quantum logic)~\cite{Abramsky2020logic} and the settings of two structural theories to understand contextuality (presented in section~\ref{CSW} and~\ref{Abramsky}), being able to connect the features and properties of these different frameworks. See also~\cite{Domenech2005} for the construction of a contextual logic.

\subsection{Is logic Empirical?}
\label{sub:neopositivism}

It may be difficult today to understand why the question of the origin of logic could seem so important in the 50's. Let us briefly recall that the Vienna Circle, which was an influential school of philosophy, had for project to give a precise and formal definition of science, notably through a careful analysis of the language, and to reject metaphysical thesis on the basis of their absence of meaning. The distinction between synthetical and analytical propositions was emphasized, and accordingly any proposition could be either shown to hold logically or to be empirically verified (a thesis which was profoundly attacked by Quine in 1951 in the famous paper\textit{Two dogmas of empiricism}~\cite{Quine1951}). It is in this context of boiling epistemological considerations that Putnam claimed to build logic on empirical considerations~\cite{putnam1969} and that Specker developed his answer to the future contingent quarrel. Karl Popper led another influential criticism, rejecting the idea that metaphysical concepts were meaningless, and stating that the science area should be delimited through the notion of falsifiability. Let us note that in this perspective, the sentence \emph{It will or it will not rain}, equivalent to \emph{There will or there will not be a sea battle tomorrow}, is introduced by Popper as a paradigmatic example of non-empirical sentence, which can not be falsified~\cite{popper2005}.

\subsection{A context of quantum foundations}
\label{sub:QLogic}
\par In 1951, Specker assisted to a seminar organised by Gonseth about quantum foundations and logic, where the work of von Neumann was extensively discussed~\cite{VonNeumann1932}. Note that von Neumann had already established a theorem about contextuality - which was false, as Bell would prove years later~\cite{Mermin1993}. It was a time of tremendous work on the questions of hidden-variables. Specker cited it as a decisive moment for his article even if he eventually published it only ten years afterwards. Among the participants of the seminar were Borel, Pauli, Destouches and Février. Destouches was a researcher in logic and philosophy who notably established a logical formalism able to state if two theories could be united or not. In 1939, Destouches wrote: “\textit{if two theories are such that no proposition has its negation in the other, there is a general theory that subsumes them, for which [...] the axioms are the logical product of all the axioms of the theory}"~\cite{Destouches1939}. But the more interesting case is the one of incompatible theories, those for which the associated models possess contradictory propositions. The only way to unify them consists in getting rid of the classical boolean rules of logic and using those invented by Reichenbach~\cite{Bitbol2001}, notably a three-valued logic~\cite{Murzi}.

\subsection{Destouches work}
According to Bitbol~\cite{webBitbol, Bitbol2001}, the pragmatic consequence of the formal system of Destouches was to prevent oneself from stating about the veracity of a proposition without taking into account the context in which it had been enunciated, and to refuse to give a meaning to a proposition if it had not been first verified that the context in which it is taken does not imply the conjunction of incompatible contexts \cite{Bitbol2016}. Destouches gave as a fundamental illustration of his work the deduction of Bohr complementarity principle within his own logical framework by unifying the incompatible particle and wave theories of electromagnetic field thanks to a change in the logical rules structure. His non-classical logic algebra was indeed logically isomorphic to quantum mechanics. Hence, a theory that unifies models with contradictory propositions must necessary take the context of experiments into account. The main ideas of this work can be found in~\cite{Destouches1948}, currently being translated in English. The reader may also consult a sum-up of the main axis of this work in~\cite{Delhotel2004}.

\subsection{The Gleason Theorem and Quantum Mechanics as a prevision theory based on contextuality}
From a philosophical perspective this indicates, according to Bitbol, that the traditional method of physicists to study objects, without a reflexive understanding on the conditions with which they acquire information about them does no longer hold~\cite{Bitbol2010}. Some researchers even suggested that this could constitute the backbone of our understanding of quantum mechanics. Within the theoretical framework that J-L. Destouches had elaborated, Paulette Février proved that the Born rule emerges as a necessary requirement for any probabilistic theory with previsions that takes measurement contexts into account~\cite{Fevrier1946}. It could be interesting to put that into perspective~\cite{BitbolConv} with the derivation of the Gleason theorem~\cite{Gleason1957} several years later, that shows that Born rule is a consequence of the usual postulate of quantum mechanics and one of non-contextuality. Another derivation of the Born rule, linked with the Gleason theorem, can be found in~\cite{Auffeves2017, Auffeves2020}, based on an ontology of quantum theory where the physical properties of systems are distributed in both the systems and the contexts in which they are embedded~\cite{Auffeves2016}. Gleason theorem had a major influence on Specker. It was actually shown that the Kochen-Specker theorem could be seen as a consequence of Gleason's: it constructs a finite set of rays on which no two-valued homomorphism exists, the existence of which is established by a simple extension to Gleason’s theorem~\cite{Hrushovski2004s, Marzlin2015}.

\par We give the formal expression of the Gleason theorem below, reproducing~\cite{GleasonTheorem}. For a discussion on the links between Gleason and KS theorem, see~\cite{Moretti2019fundamental}.

\begin{definition}{\textbf{Finite additive measure}}
Let $\rho: \mathcal{P} \to [0, 1]$ such that for every finite family $\{ P_1, ..., P_n: P_i \in \mathcal{P} \}$ of pairwise orthogonal projections we have $\rho(\sum_{i=1}^n P_i) = \sum_{i=1}^n \rho(P_i)$ , then $\rho$ is a finitely additive measure on $\mathcal{P}$.
\end{definition}

\begin{theorem}{\textbf{Gleason's theorem}}
Let $\mathcal{H}$ be a Hilbert space. If $\dim(\mathcal{H}) \neq 2$, then each finitely additive measure on $\mathcal{P}$ can be uniquely extended to a state on $\mathcal{B}(\mathcal{H})$. Conversely the restriction of every state to $\mathcal{P}$ is a finitely additive measure on $\mathcal{P}$.
\end{theorem}

\subsection{A tale of contextuality...}
\label{sub:tale}
In \textit{The logic of non-simultaneously decidable propositions}, Specker illustrated the logical propositions he questioned with a parabola, of which I give here a shorter and personal version. The reader is invited to consult the original one~\cite{Specker1960}. Liang, Wiseman and Spekkens have also extensively studied this example, using it to introduce several key concepts and results of contextuality~\cite{Liang2011}.

\par 
\textit{A long time ago, in the Assyrian province of Ninive, was a very wise man, whose ability to understand mathematics and physics had turned himself into a magician in the eyes of his people: he was able to predict the solar and lunar eclipses without failures. Proud of his achievements and eager to share his knowledge, he opened a course for graduate Assyrians. Unfortunately, the students were as keen on watching astrological phenomena as they were lazy on calculating trajectories of stars. It turned out, however, that, they were very much interested in another aspect of his life: his daughter. When she reached the marriageable age, he received tons of marriage proposals, so much so that he invented a test to find the suitable husband. He presented three boxes, and guaranteed that, among the three, at least one contained a gem and one was empty, and asked suitors to indicate two that were full or two that were empty. It happened that every time they opened two boxes, one was full and the other empty. At first, the daughter of the prophet followed the game from distance, but one day she agreed with a student she loved that they should be married. She went to her father, and she indicated two boxes, saying that one was full and the other empty. She opened the boxes and her prediction was confirmed. Her father protested weakly, saying that she would have opened two other boxes, had she respected the rules of the game, but when she tried to open the third one, a magic force prevented her from doing so. He then declared that her prediction was valid, and went on to meditate on the relevance of mathematical tests with regards to the happiness of children.}

\subsection{...and its logical implications}
\label{sub:explanation}
Let us present a logical interpretation of this tale, according to Specker~\cite{Specker1960}. We note the six propositions “The first/second/third box is full/empty" with the symbols $A_i$ or $A_i^*$. For $ \{i,j\} \in \{1,2,3\}, i \ne j$, $A_i$ indicates that the i-th box contains a gem, $A_i^*$ that it is empty. The enchantment of the gem makes it respect the following conditions: 
\begin{align}
    &A_i \rightarrow A_j^* \\
    &A_i^* \rightarrow A_j \\
    &A_i \rightarrow A_i \\
    &A_i^* \rightarrow A_i^*.
\end{align}
The first condition explains the failure of all the students, the second one that the prophet was not a cheater (he did not teleport the gem). Now, the trick is that this yields 
\begin{equation}
    A_1 \rightarrow A_2^* \rightarrow A_3
\end{equation}
...but of course also
\begin{equation}
    A_1 \rightarrow A_3^* !
\end{equation} Hence, these logical conditions can hold only if one of the box is not opened. 
\par We can interpret it as saying that the joint measurement of the three boxes is impossible. If the measurements (opening a box) are to reveal a pre-determined value (the presence or the absence of a gem), the measurement outcome must depend on the context (the presence or the absence of the gem when we have open the first box). The analogy with quantum mechanics is limited by the following property: in Specker's parabola, the fact that each pair of observables is jointly measurable does not imply that all observables are compatible. This is the case in quantum mechanics: if we have a set of three observables, $A,B,C$, such that $[A,B] = 0$, $[B,C] = 0$ and $[A,C] = 0$, then the three observables are jointly measurable, and we do not need anymore the hypothesis of contextuality to explain the results of measurements. 

\par In 1961, Specker presented this article in a seminar in which he met Simon Kochen, a mathematician. Their collaboration led to a quantum mechanics formulation of this theorem that was to be known under the name of Kochen-Specker theorem. According to the formulation of Spekkens~\cite{Dourdent2018}, Specker had turned Thomas Aquinas question into the following one: “Could [God] know the result that would have been obtained, had another quantum measurement been performed than the one actually performed, and this without creating contradictions?".


\section{Mathematical proofs of contextuality}
\label{MathematicalProofs}
The first proof of contextuality of quantum mechanics was the Kochen-Specker Bell theorem, that we present here after. New proofs were derived afterwards, Clifton~\cite{Clifton1992} for instance insisting on simplifying the geometrical arguments. In the 90's, Peres demonstrated the contextuality of Quantum Mechanics in a Hilbert space of dimension four with a particular state, Mermin \cite{Mermin1993} then transformed his argument into a state independent proof of contextuality. As links between non-locality and contextuality began to be assessed, research began to be undertaken with the objective of testing experimentally this property. To this end, non-contextuality inequalities were derived in the 2000's, first by Klyachko, Can, Binicio\u{g}lu and Shumovsky~\cite{Klyachko2008}, then by Cabello~\cite{Cabello2008} following the proof of Peres and Mermin. It has been proven that any proof of the Kochen–Specker theorem can always be converted to a state-independent noncontextuality inequality~\cite{Yu2015}. For a full review of the proofs, see~\cite{Budroni2021}. Recently, the Kochen-Specker has been shown to follow from the Burnside theorem, applied on non-commutative algebra~\cite{Bolotin2018}. 
\par We restrict this section to a pedagogical presentation of the KS theorem so that the reader can seize the mathematical important ideas, before deriving the KCBS inequality and detailing the proof of the Peres-Mermin square. The KCBS inequality will be used regularly to present the different structural theories of contextuality, while the Peres-Mermin square is merely at the heart of our discussion on the loopholes of contextuality experimental proofs. 

\subsection{The Kochen-Specker Bell theorem}
\label{sub:KS}
Let us note $A$ and $B$ two commuting observables. There exists a common basis $\Pi_k$ of projectors, such that:
\begin{equation}
    A = \sum_k a_k \Pi_k
\end{equation}
and
\begin{equation}
    B = \sum_k b_k \Pi_k.
\end{equation}
This common basis makes it possible to measure simultaneously $A$ and $B$.
\begin{equation}
    AB = \sum_k a_k b_k \Pi_k
\end{equation}
and
\begin{equation}
    A+B = \sum_k (a_k+b_k) \Pi_k.
\end{equation}
Now, let us assume the hypothesis called \textit{Outcome Determinism} (O.D.): the fact that the values of an observable preexist to any measurement. We can associate to any observable an outcome, according to 
\begin{equation}
    v(A) = \sum_k a_k v(\Pi_k)
\end{equation}
and $v$ has a finite set of possible results. We can restrict it without loss of generality to $0$ and $1$:
\begin{equation}
    v(A) \in \{0,1\}.
\end{equation}
According to the \textit{Measurement Non-Contextuality} (MNC) hypothesis, the outcome of an observable does not depend on the context in which it is being measured. Thus,  
\begin{equation}
    v(AB) = v(A) v(B)
\end{equation}
and
\begin{equation}
    v(A+B) = v(A)+v(B)
\end{equation}
The Kochen-Specker (KS) theorem proved that in any Hilbert space whose dimension is greater than 3, the predictions of quantum mechanics (QM) are not consistent with the two previous hypothesis. To do so, it used the squared of the components of particle of spin-1 $S$ in orthogonal directions $u,v,w$ as commuting observables, and proved that the equality 
\begin{equation}
    S_u^2 + S_v^2 + S_w^2 = 2
\end{equation}
\cite{KochenSpecker1967, Mermin1993} can not be fulfilled by a deterministic hidden variable model for a particular set of directions, whereas a quantum state can. The full proof is rather complex as it involves 117 directions. The logical conclusion of this theorem can be written:
\begin{equation}
    QM \wedge OD \wedge MNC = \emptyset
\end{equation}
where $\wedge$ is the logical intersection symbol. This theorem does not rule out noncontextuality but it imposes to choose between the validity of the three hypothesis. After having expressed the logical contradiction within the framework of quantum mechanics, the next step was to derive inequalities able to discriminate between hidden variable noncontextual models and quantum mechanics.

\subsection{KCBS inequality}
\label{sub:KCBS}
The first proposal of that sort was made by Klyachko, Can, Binicio\u{g}l{u} and Shumovksky, and is called the KCBS inequality~\cite{Klyachko2008}. It uses  five  different observables $A_i (\{ i\in\{1:5\})$, with binary outcomes $\{ \pm 1 \}$~\cite{Klyachko2008, Jerger2016}. The test involves measuring the five pairs of observables, called measurement contexts, $\{A_1,A_2\}$, $\{A_2,A_3\}$, $\{A_3,A_4\}$,$\{A_4,A_5\}$ and $\{A_5,A_1\}$, chosen such that each observable is measured in two different contexts. Noncontextual HV models predict that the total observable correlations for outcome pairs are lower bounded by -3.
\begin{equation}
    \moy{A_1 A_2} + \moy{A_2 A_3} + \moy{A_3 A_4} + \moy{A_4 A_5} + \moy{A_5 A_1} \geq - 3.
    \label{eq:KCBS}
\end{equation}
We can see it graphically by representing this system by a pentagon, for which each vertex is the measurement result of an observable, and an edge a context. Indeed, the only way to minimise the above sum is to alternate the sign of the outcome as much as possible.

\begin{figure}
\centering
  \begin{tikzpicture}[
  roundnode/.style={circle, draw=green!60, fill=green!5, thick},
squarednode/.style={diamond, draw=red!70, fill=red!5,  thick},
  ]
    \draw [thick] (-1,0)--(1,0) ;
    \draw [thick] (1,0)--(1.62,1.9) ;
    \draw [thick] (1.62,1.9)--(0,3.08) ;
    \draw [thick] (0,3.08)--(-1.62,1.9) ;
    \draw [thick] (-1.62,1.9)--(-1,0) ;
    \draw (-1,0) node[above right]{$A_1$};
    \draw (1,0) node[above left]{$A_2$};
    \draw (1.62,1.9) node[right]{$A_3$};
    \draw (0,3.08) node[above] {$A_4$};
    \draw (-1.62,1.9) node[left]{$A_5$};
    \draw (-1,0) node[squarednode] [below]{$-1$};
    \draw (1,0) node[roundnode] [below]{$+1$};
    \draw (1.62,1.9) node[roundnode] [left]{$ -1 $};
    \draw (0,3.08) node[squarednode] [right] {$ +1 $};
    \draw (-1.62,1.9) node[squarednode] [above]{$ -1 $};
  \end{tikzpicture}
  \caption{The KCBS pentagram | We display a specified pre-assignment outcome that maximises the inequality~\ref{eq:KCBS}.}
  \label{fig:PentagonKCBS1D}
\end{figure}
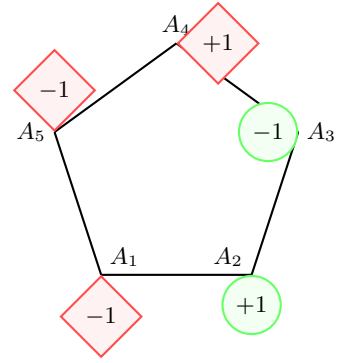

This inequality can be violated in quantum mechanics. We consider five dichotomic observables $A_i = 2 \ket{l_i} \bra{l_i} - 1$. They can be described by a pair of projectors ${\ket{l_i}\bra{l_i},\mathds{1} - \ket{l_i}\bra{l_i}}$ associated with outcomes  $\{ \pm 1 \}$.  The  states  connected  by  edges  of the pentagram are orthogonal, ensuring that the corresponding observables, $A_i$ and $A_{i+1}$ are compatible. The projectors are, explicitly, in ket notation~\cite{Dourdent2018, Howard2013}:
\begin{align*}
    &\ket{l_1} = \frac{1}{\sqrt{2}} \ket{1,0,\sqrt{\cos{(\pi/5)}}}\\
    &\ket{l_2} = \frac{1}{\sqrt{3}} \ket{\cos{(4\pi/5)},\sin{(4 \pi/5)},\sqrt{\cos{(\pi/5)}}}\\
    &\ket{l_3} = \frac{1}{\sqrt{3}} \ket{\cos{(2\pi/5)},- \sin{(2 \pi/5)},\sqrt{\cos{(\pi/5)}}}\\
    &\ket{l_4} = \frac{1}{\sqrt{3}} \ket{\cos{(4\pi/5)},- \sin{(4 \pi/5)},\sqrt{\cos{(\pi/5)}}}\\
    &\ket{l_5} = \frac{1}{\sqrt{3}} \ket{\cos{(2\pi/5)},- \sin{(2 \pi/5)},\sqrt{\cos{(\pi/5)}}}\\
\end{align*}

In this case, equation \ref{eq:KCBS}  yields $5-4 \sqrt{5} \approx -3.9$ for the qutrit $\ket{\psi} = \ket{0,0,1}$ This is the maximum quantum violation of inequality \ref{eq:KCBS}~\cite{Cabello2013}. We note that this proof is state-dependent, which was not the case in the earlier KS proofs. This is however commonly known in the field of nonlocal Bell inequalities~\cite{MansfieldConv}.

The same inequality can be adapted to yield 
\begin{equation}
    \moy{A_1 A_2} + \moy{A_2 A_3} + \moy{A_3 A_4} + \moy{A_4 A_5} + \moy{A_5 A_1} \leq 2
    \label{eq:KCBS2}
\end{equation}
when the possible outcomes of $A_i$ are $\{0,1\}$. Both perspectives are used in the literature.

\subsection{Peres-Mermin Square}
\label{sub:PM}
Peres and Mermin \cite{Mermin1993} later proposed a particularly simple proof of contextuality called the Peres-Mermin Square, which had the important advantage of being state-independent. It was later turned into an inequality by Cabello~\cite{Cabello2008}. Consider a collection of $9$ dichotomic observables $\{A_{ij}\}$, chosen such that the observables are compatible when they share a common subscript. Hence, it is possible to measure the product of observables inside a column or a row (which constitutes a context), as they appear in Table~\ref{tab:PMA}, and to form the quantity:
\begin{align}
\moy{S} &= \moy{A_{11}A_{12}A_{13}} + \moy{A_{21}A_{22}A_{23}} +  \moy{A_{31}A_{32}A_{33}} \nonumber \\  &+\moy{A_{11}A_{21}A_{31}} +  \moy{A_{12}A_{22}A_{32}} - \moy{A_{13}A_{23}A_{33}}.
\end{align}

\begin{table}
\centering
\begin{tabular}{|c|c|c|c|c|c|c| }
\hline $A_{jk}$     & k = 1  &     k = 2    &    k = 3 \\
\hline j = 1      &  $A_{11}$ & $ A_{12} $ & $ A_{13} $ \\
\hline j = 2     &  $A_{21}$ & $ A_{22} $ & $ A_{23} $  \\
\hline j = 3      &  $A_{31}$ & $ A_{32} $ & $ A_{33} $ \\
\hline
\end{tabular}
\caption{The Peres-Mermin square for classical binary observables}
\label{tab:PMA}
\end{table}

The original proof of Peres-Mermin considers a hidden-variable model, where the values of $\{A_{ij}\}$ are determined by a $\lambda$. In this formalism, $A_{ij}(\lambda) \in \{\pm 1\}$, according to the \textit{outcome determinism} hypothesis, and $v(A_{i1}A_{i2}A_{i3}) = v(A_{i1}) v(A_{i2}) v(A_{i3})$, where $v(A_{ij})$ is the value of observable $A_{ij}$, according to the \textit{measurement noncontextuality} hypothesis. In this case, all the $2^9$ possible configurations of outcomes of observables $\{A_{ij}\}$ give 
\begin{equation}
    \moy{S} \leq 4.
    \label{eq:PM}
\end{equation}
On the contrary, in quantum mechanics, the dichotomic observables are given by hermitian operators with a binary spectrum. We can for instance consider the case of observables made out of tensorial product of two Pauli operators. They are chosen such that the operators within a context, \textit{i.e.} within a row or a column commute. The table of Peres-Mermin for these operators is given in \ref{tab:PMQM}. Because the product of operators in each row and column is equal to $\mathds{1}$, except for the last row when it is equal to $-\mathds{1}$, $\moy{S}_{QM} = 6$, for any quantum state and thus the measurement outcome predicted by quantum mechanics can not be reproduced by any KS-contextual model. 

\begin{table}
\centering
\begin{tabular}{|c|c|c|c|c|c|c| }
\hline $A_{jk}$     & k = 1  &     k = 2    &    k = 3 \\
\hline j = 1      &$  {\sigma_{x}} \otimes \mathds{1} $&  $\mathds{1}  \otimes {\sigma_{x}} $ &  ${\sigma_{x}} \otimes {\sigma_{x}} $ \\
\hline j = 2      & $\mathds{1}  \otimes {\sigma_{z}} $ & ${\sigma_{z}} \otimes \mathds{1}  $&   ${\sigma_{z}} \otimes {\sigma_{z}} $ \\
\hline j = 3      & ${\sigma_{x}} \otimes {\sigma_{z}} $&  ${\sigma_{z}}  \otimes {\sigma_{x}} $ &  ${\sigma_{y}} \otimes {\sigma_{y}} $ \\
\hline
\end{tabular}
\caption{The Peres-Mermin square for Pauli Observables}
\label{tab:PMQM}
\end{table}


\clearpage
\begin{center}
\textbf{\textsc{\Large The structural theories}}
\end{center}
We present in this part the main structural theories that were established to develop the notion of contextuality. The last part presented the first proofs of contextuality that were discovered when the field was emerging. This one focus on a systematic approach to understand contextuality. We begin with the graph-theoretic approach in section~\ref{CSW}, which is first and foremost a generalisation of the previously derived inequalities. A larger perspective is given by the sheaf-theoretic approach, presented in section~\ref{Abramsky}. It is based on probabilistic and category models, and leads notably to a hierarchy of contextuality that paves the way towards a resource theory for quantum computing. We present the hypergraph theoretic approach that encompasses both of them in section~\ref{ALFS}. A major conceptual step had been established by Spekkens before, thanks to the presentation of an operational theory of contextuality that enlarges its scope beyond Quantum Mechanics. The shadow of this approach looms in all the others, but we chose to present it only in section~\ref{SpekkensContextuality} so that its reinterpretation of the previous approaches is clearer and more motivated. In the next and last part, we will use Spekkens framework to understand the issues that arise when one wishes to turn the proofs of contextuality considered in this one into experimentally robust ones. This is why we present the main ingredients of the last structural theory, Contextuality-by-Default (CbD), in section~\ref{Cbd}, as we will use it to extend propositions of the sheaf-theoretic approach into an adequate answer to Spekkens. Finally, we sketch a comparison between the theories in section~\ref{comparisons}, and present some connections between theories inside a table.

\section{The graph-theoretic approach of Cabello, Severini and Winter (CSW)}
\label{CSW}
Once the KCBS, or the Peres-Mermin square based inequality (See sections~\ref{sub:KCBS} and~\ref{sub:PM}) were established, Cabello, Severini and Winter (CSW) proposed a general framework for contextuality inside which they could obtain maximal and general bounds for non-contextuality inequalities \cite{CSW2014, Cabello2013}. This approach analyses to what extent a joint probability distribution can reproduce the marginals of quantum theory in the case of contextual correlations. These correlations are mapped inside a graph for which noncontextual theories, quantum theory and general probabilistic theories yield different sets of probabilities. 

\subsection{General definitions}
\label{sub:CSWdefinitions}
Let us consider a correlation experiment. Whatever the apparatus, particles or physical quantities involved, we  keep only three elements: the tests (a set of input), the outcomes and their probabilities. The recording of an outcome, given a test, is called an \textit{event}. Two events $e_i$ and $e_j$ are \textit{equivalent} if and only if they happen with the same probability, and \textit{exclusive} if they can not happen at the same time. The formal definition of exclusivity is the existence of two jointly measurable observables able to distinguish between them. We associate to any experiment a graph $\mathcal{G}$, called an \textit{exclusivity graph of the experiment}, for which the vertices correspond to events and adjacent vertices are pairs of exclusive events. 

\subsection{The case of KCBS as a paradigmatic example}
\label{sub:KCBSGraph}
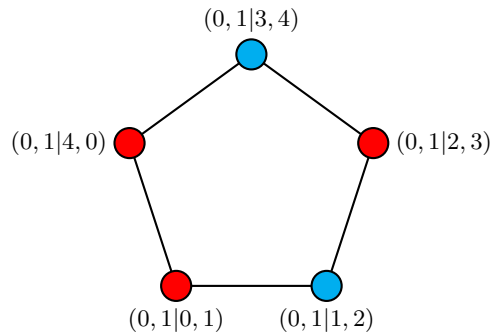
\begin{figure}
	\centering
	
\begin{tikzpicture}[scale=1]

\draw[thick] (-1,0) -- (1,0) -- (1.62, 1.90) -- (0,3.08) -- (-1.62,1.90) -- cycle;

\fill[red](-1,0) circle (0.2);
\fill[cyan] (1, 0) circle (0.2);
\fill[red] (1.62, 1.90) circle (0.2);
\fill[cyan] (0, 3.08) circle (0.2);
\fill[red] (-1.62, 1.90) circle (0.2);

\draw[thick] (-1,0) circle (0.2) node[below=5pt]{$(0,1|0,1)$};
\draw[thick] (1, 0) circle (0.2) node[below=5pt]{$(0,1|1,2)$};
\draw[thick] (1.62, 1.90) circle (0.2) node[right=5pt]{$(0,1|2,3)$};
\draw[thick] (0, 3.08) circle (0.2) node[above=5pt]{$(0,1|3,4)$};
\draw[thick] (-1.62, 1.90) circle (0.2) node[left=5pt]{$(0,1|4,0)$};
\end{tikzpicture}
  \caption{Exclusivity graph of the KCBS experiment. We specified the possible measurements results according to the inputs, just like~\cite{CSW2014}. We displayed in blue the value $1$ and in red the value $0$, evidencing the classical bound.}
  \label{fig:PentagonKCBS}
\end{figure}

Let us take the example of the KCBS inequality defined in Equation~\ref{eq:KCBS2}, section \ref{sub:KCBS}. It has five tests, noted $P_i$, with $0 \leq i \leq 5 $ and two possible outcomes, $0$ and $1$. Its possible events are accordingly “\textit{obtaining $0$ or $1$ after a test $P_i$}", which is noted: $\{(0|P_i), (1|P_i)\}$. The experiment consists in performing five pairs of tests on systems in the same quantum state, and to compute the sum of the outcomes, according to Equation~\ref{eq:KCBS2Proba}:
\begin{equation}
    S_{\text{KCBS}} = \sum_{i=0}^4 P(0,1|i,i+1).
    \label{eq:KCBS2Proba}
\end{equation}
where $P(0,1|i,i+1)$ is the probability to obtain the outcome $0$ or $1$ as the product of the outcome measurements of the correlated pair of events $i, i+1$.
We already know that for a classical model, $S_{\text{KCBS}}$ is bounded by $2$. If we represent the exclusivity graph of the KCBS experiment, we obtain a pentagon as given in Figure~\ref{fig:PentagonKCBS}.

This classical bound is called the \textit{Independence number} of the graph, and it is noted ${\alpha}$. We give below a formal definition of this concept. Intuitively this number appears on the graph as the maximum possible number of vertices which are not directly connected by any edge. This bound is violated by quantum theory, for which $S = \sqrt{5} \ge 2$.

\begin{definition}{\textbf{Independence number}}
The Independence number of a graph is the cardinality of the largest independent vertex set. 
\end{definition}
\begin{definition}{\textbf{Independent vertex set}}
An independent vertex set of a graph $G$ is a subset of the vertices such that no two vertices in the subset represent an edge of $G$.
\end{definition}

\subsection{Generalisation}

The KCBS inequality is a particular example of noncontextuality inequalities. In general, the probability of events may be weighted, so that the hidden-variable side of these inequalities becomes:
\begin{equation}
    S = \sum_i w_i P(e_i) 
\end{equation}
where the $w_i \geq 0$ are the weights of each outcome. In KCBS, all $w_i = 1$.

 We consequently define a vertex-weighted graph $(G,w)$ as a graph $G$ with a set $V$ of vertices weighted by a positive real function $w$. We can now present the powerful result of~\cite{CSW2014}:
 
 \begin{proposition}{\textbf{Bounds on noncontextual inequalities from graph theoretic approach}}
Let $S$ be a noncontextual inequality. The the following bounds apply: 
\begin{equation}
    S \overset{NCHV}{\leq} \alpha (G,w) \overset{Q}{\leq} \mathcal{V}(G,w)
\end{equation}
\end{proposition}
where $\alpha (G,w)$ is the independence number of $G,w$, and $\mathcal{V}(G)$ is the Lov\'asz function of the graph~\cite{Lovasz1979}. These concepts are defined below.

\begin{definition}{\textbf{Independence number of a graph}}
The \textit{independence number} of a weighted graph $(G,w)$ is the maximum value of the sum $\sum_{i \in I} w_i$ where $I$ is any independent set.
\end{definition}

\begin{definition}{\textbf{Orthonormal representation of a graph}}
The set of $\ket{\phi_i}$ is called an \textit{orthonormal representation} of $G$ when each $\ket{\phi_i}$ is a unitary vector of $\mathbf{R}^d$ associated to a vertex $v_i \in V$ such that $\braket{\phi_i | \phi_j} = 0$ for two nonadjacent vertices $v_i$ and $v_j$. The dimension $d$ is arbitrary.
\end{definition}

\begin{definition}{\textbf{Lov\'asz function}}
The Lov\'asz function of $(G,w)$ is the maximum value of 
\begin{equation}
    \sum_{i \in V} w_i |\braket{\phi_i|\psi}|^2 
\end{equation}
over any orthonormal representation of $\bar{G}$,  where $\ket{\psi} \in \mathbf{R}^d$ is the state of the system, and $\bar{G}$ is the complement of $G$
\end{definition}

\begin{definition}{\textbf{Complement of a graph}}
The complement $\bar{G}$ of a graph $G$ is a graph on the same vertices such that two distinct vertices of $\bar{G}$ are adjacent if and only if they are not adjacent in $G$.
\end{definition}

\par The most remarkable feature of this approach is to show that $\mathcal{V}(G)$, which is a feature of graph theory, is the maximum value of $S$ given by the quantum theory. It can be computed for any graph in polynomial time at a given precision as it is the optimal solution of a SDP program~\cite{Giandomenico2013}. Further details on why this is possible are presented in~\cite{CSW2014}.

\subsection{Link with Bell inequalities}
\label{sub:BellCSW}
As we mentioned above, Bell inequalities share a lot of properties with noncontextual inequalities. They are also derived from deterministic hidden-variable models applied on correlation experiments. The graph theoretic framework of CSW also works for Bell inequalities and relates the Tsirelson bound to graph properties. The analogous of the KCBS inequality, the so-called CHSH inequality, is explained in the same manner. Besides, it has been shown that under certain hypothesis, notably ideal measurements, Bell non-locality and KS-contextuality are equivalent inside quantum theory. This means that every quantum violation of a Bell inequality can be transformed into a quantum violation of KS-noncontextual inequality~\cite{Suarez2017}, and \textit{vice versa}~\cite{Cabello2019}.


\section{A sheaf-category theoretic approach to contextuality}
\label{Abramsky}
The framework of graph-theory had notably enabled CSW to derive general inequalities of noncontextuality, to express their properties and to shed light on the links with Bell inequalities. The framework of sheaf-theory allowed Pr. Abramsky and Brandenburger to give an even larger perspective~\cite{Abramsky2011}. Category theory is the branch of mathematics that formalises mathematical structures and study their connections. Inside it, sheaf theory has been efficient at explaining the links between local and global solutions to a given problem. This structure subsumes contextuality and non-locality in a unified way, and shows that they can be characterised as “obstruction to the existence of global" solutions~\cite{Abramsky2011, Abramsky2017}. A hierarchy of contextuality is also derived. The notion of compatibility is reinterpretated within this framework, and this formalism paves the way towards quantifying the amount of contextuality of a scenario~\cite{Mansfield2017conf, Abramsky2017}, which will be fruitfully summoned in the last parts of this work. This approach was later extended in several dimensions, from continuous variables~\cite{Mansfield2019} to contextuality witness \cite{Abramsky2017conf} and the specification of states for which contextuality can happen~\cite{Mansfield2019}.

We introduce the subtle concepts of this approach with a graphic representation of the three types of contextuality that it hierarchises, that was initially introduced in~\cite{Abramsky2015}. Then, we introduce the mathematical structures from which they appear.
\subsection{Pictorial representation}
\label{sub:Pictorial}
We consider a correlation experiment where two operators, Alice and Bob can measure respectively two bivalued observables $\{A,A'\}$ and $\{B,B'\}$, one at a time. This is a typical CHSH scenario. The four possible outcomes form a set $\mathcal{O} = \{00, 01, 10, 11 \}$. We form a probability table where the measurements performed are linked with their probabilities. This is given in table ~\ref{tab:SheafProba}, where we indicate that the probability of obtaining $00$ when $A$ and $B$ are measured is $1/2$, of obtaining $10$ when $A'$ and $B'$ are measured is $3/8$, etc...
\begin{table}[h!]
\centering
\begin{tabular}{c c c c c}
            & 00  & 01  & 10 & 11 \\
\hline A B  & 1/2 & 0 & 0 & 1/2  \\
\hline A B' & 3/8 & 1/8 & 1/8 & 3/8  \\
\hline A' B & 3/8 & 1/8 & 1/8 & 3/8  \\
\hline A' B'& 1/8 & 3/8 & 3/8 & 1/8  \\
\end{tabular}
\caption{The probability table of the CHSH experiment: local projective measurements equatorial at angles $0$ and $\pi/3$ on the Bell state $\ket{\phi^+} = \frac{1}{2} \left( \ket{00} + \ket{11} \right)$~\cite{Abramsky2017}.}
\label{tab:SheafProba}
\end{table}

From this table, we derive a new one, called a possibility table, where we only keep the information that a correlation is possible, when its probability is above 0, or impossible, when it is equal to 0. A line represents the results of jointly measurable observables inside a context. The rows correspond to the support of some unspecified probability distributions. 

\par When the table is complete, we can represent its topological appearance by constructing a \textit{contextuality bundle} diagram. To do so, we represent all the observables as vertices, and we connect compatible ones with edges. We associate to every observable its possible measurement outcomes: this is called a \textit{fibre}. We connect with edges the possibly jointly measured outcomes, those tagged with a $\checkmark$ in the table, and do not connect those tagged with a $\xmark$. This is done in Tables~\ref{tab:SheafCHSH},~\ref{tab:SheafLogical} and~\ref{tab:SheafKCBS}. On this representation, contextuality will be visualised by the presence among these edges of closed paths traversing all the fibres univocally \textit{i.e.} exactly once~\cite{Abramsky2017}. It corresponds to a global assignment to local values consistent with the model.

\subsubsection{Probabilistic contextuality}
\label{subsub:ProbaContex}
We present the possibility table of the CHSH scenario in Table~\ref{tab:SheafCHSH}.

\begin{table}
\centering
\begin{tabular}{c c c c c}
            & 00  & 01  & 10 & 11 \\
\hline A B  & $\checkmark$ & $\xmark$ & $\xmark$ & $\checkmark$  \\
\hline A B' & $\checkmark$ & $\checkmark$ & $\checkmark$ & $\checkmark$  \\
\hline A' B & $\checkmark$ & $\checkmark$ & $\checkmark$ & $\checkmark$  \\
\hline A' B'& $\checkmark$ & $\checkmark$ & $\checkmark$ & $\checkmark$  \\
\end{tabular}
\caption{The possibility table of the CHSH experiment.}
\label{tab:SheafCHSH}
\end{table}

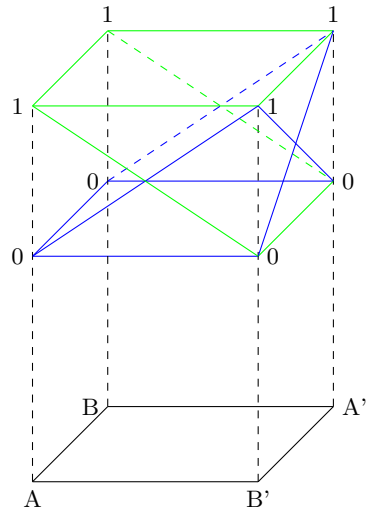
\begin{figure}[h!]
\centering
  \begin{tikzpicture}
    \draw(1,0)--(4,0) ;
    \draw(2,1)--(5,1) ;
    \draw(1,0)--(2,1) ;
    \draw(4,0)--(5,1) ;
    \draw (1,0) node[below]{A};
    \draw (4,0) node[below]{B'};
    \draw (2,1) node[left]{B};
    \draw (5,1) node[right] {A'};
    \draw[blue](1,3)--(4,3) ;
    \draw[blue](2,4)--(5,4) ;
    \draw[blue](1,3)--(2,4) ;
    \draw[green](4,3)--(5,4) ;
    \draw (1,3) node[left]{0};
    \draw (4,3) node[right]{0};
    \draw (2,4) node[left]{0};
    \draw (5,4) node[right] {0};
    \draw[green](1,5)--(4,5) ;
    \draw[green](2,6)--(5,6) ;
    \draw[green](1,5)--(2,6) ;
    \draw[green](4,5)--(5,6) ;
    \draw (1,5) node[left]{1};
    \draw (4,5) node[right]{1};
    \draw (2,6) node[above]{1};
    \draw (5,6) node[above] {1};
    \draw[dashed] (1,0)--(1,5);
    \draw[dashed] (4,0)--(4,5);
    \draw[dashed] (2,1)--(2,6);
    \draw[dashed] (5,1)--(5,6);
    \draw[dashed,green](2,6)--(5,4) ; 
    \draw[dashed, blue](2,4)--(5,6) ; 
    \draw[blue](4,5)--(5,4) ; 
    \draw[blue](4,3)--(5,6) ; 
    \draw[blue](4,5)--(1,3) ; 
    \draw[green](4,3)--(1,5) ; 
  \end{tikzpicture}
  \caption{Bundle diagram of the CHSH experiment. The green and blue coloured edges correspond to possible correlations. Besides, the green edges form two particular closed univocal path.}
  \label{fig:BundleCHSH}
\end{figure}

We see on its bundle, depicted in blue colour, Figure~\ref{fig:BundleCHSH}, that except for the couple $A$ and $B$, for which only two edges appear between $0$ and $0$, and $1$ and $1$, four edges are drawn, depicting the fact that all these correlations are possible.

It is always possible, starting from one edge (one local assignment), to draw a closed univocal path that will go back to the first node. We highlight the presence of some of these paths in green colour. This scenario is at the lowest level of contextuality, one for which contextuality arises from probability distributions only.

\subsubsection{Logical contextuality}
\label{subsub:LogicalContex}
\begin{figure}[h!]
\centering
  \begin{tikzpicture}
    \draw(1,0)--(4,0) ;
    \draw(2,1)--(5,1) ;
    \draw(1,0)--(2,1) ;
    \draw(4,0)--(5,1) ;
    \draw (1,0) node[below]{A};
    \draw (4,0) node[below]{B'};
    \draw (2,1) node[left]{B};
    \draw (5,1) node[right] {A'};
    \draw[red](1,3)--(2,4) ;
    \draw[green](4,3)--(5,4) ;
    \draw (1,3) node[left]{0};
    \draw (4,3) node[right]{0};
    \draw (2,4) node[left]{0};
    \draw (5,4) node[right] {0};
    \draw[green](1,5)--(4,5) ;
    \draw[blue](2,6)--(5,6) ;
    \draw[blue](1,5)--(2,6) ;
    \draw (1,5) node[left]{1};
    \draw (4,5) node[right]{1};
    \draw (2,6) node[above]{1};
    \draw (5,6) node[above] {1};
    \draw[dashed] (1,0)--(1,5);
    \draw[dashed] (4,0)--(4,5);
    \draw[dashed] (2,1)--(2,6);
    \draw[dashed] (5,1)--(5,6);
    \draw[dashed,green](2,6)--(5,4) ; 
    \draw[dashed, blue](2,4)--(5,6) ; 
    \draw[blue](4,5)--(5,4) ; 
    \draw[blue](4,3)--(5,6) ; 
    \draw[blue](4,5)--(1,3) ; 
    \draw[green](4,3)--(1,5) ; 
  \end{tikzpicture}
  \caption{Bundle diagram of the Hardy paradox. The green and blue coloured edges correspond to possible correlations. We highlight in red an impossible correlation that prevents the existence of a particular global univocal path.}
  \label{fig:BundleHardy}
\end{figure}
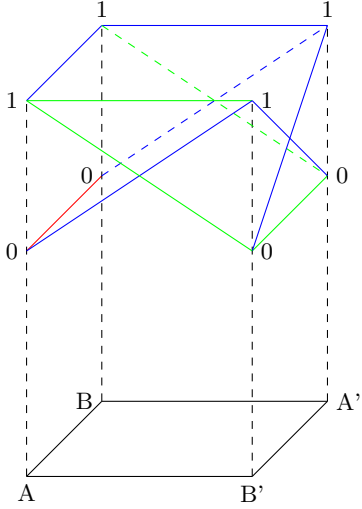

In contrast with the previous section, we give an example of a scenario, for which this is not always possible to draw an univocal closed path. To do so, we present the possibility table of the Hardy paradox, a scenario of quantum non-locality, in Table~\ref{tab:SheafLogical} and its associated bundle diagram in Figure~\ref{fig:BundleHardy}. Hardy's paradox~\cite{Hardy1992, Hardy1993} is a thought experiment in which a particle and its antiparticle interact without annihilating each other. Among the works that demonstrate Bell theorem and the contextuality of quantum mechanics without implying inequalities~\cite{Greenberger1990}, it is considered as a paradigmatic example~\cite{Silva2017, Yokota2009}.

\begin{table}
\centering
\begin{tabular}{c c c c c}
            & 00  & 01  & 10 & 11 \\
\hline A B  & $\checkmark$ & $\checkmark$ & $\checkmark$ & $\checkmark$  \\
\hline A B' & $\xmark$ & $\checkmark$ & $\checkmark$ & $\checkmark$  \\
\hline A' B & $\xmark$ & $\checkmark$ & $\checkmark$ & $\checkmark$  \\
\hline A' B'& $\checkmark$ & $\checkmark$ & $\checkmark$ & $\xmark$  \\
\end{tabular}
\caption{The possibility table of the Hardy Paradox.}
\label{tab:SheafLogical}
\end{table}

There exist some local assignments that are globally consistent with the model, for instance $\{A=1, B=1, A'=0, B'=0\}$ which is depicted in green. On the other hand, if we start with the local assignment $\{A=0, B=0\}$, depicted in red in the diagram, we can not find a global path that goes univocally through the available edges and includes it. This local assignment has no globally coherent extension. This is characteristic of the second level of the hierarchy of contextuality, and it is called \textit{logical contextuality}.

\subsubsection{Strong contextuality}
\label{subsub:StrongContex}
\begin{figure}[h!]
\centering
  \begin{tikzpicture}
    \draw (2,0) node[below]{$A_1$};
    \draw (4,2) node[right]{$A_2$};
    \draw (3,4) node[right]{$A_3$};
    \draw (1,4) node[left]{$A_4$};
    \draw (0,2) node[left]{$A_5$};
    \draw (2,0)--(4,2);
    \draw (4,2)--(3,4);
    \draw (3,4)--(1,4);
    \draw (1,4)--(0,2);
    \draw (0,2)--(2,0);
    \draw (2,7) node[below right]{$0$};
    \draw (4,9) node[right]{$0$};
    \draw (3,11) node[right]{$0$};
    \draw (1,11) node[left]{$0$};
    \draw (0,9) node[left]{$0$};
    \draw [dashed] (2,7)--(4,9);
    \draw [dashed] (4,9)--(3,11);
    \draw [dashed] (3,11)--(1,11);
    \draw [dashed] (1,11)--(0,9);
    \draw [dashed] (0,9)--(2,7); 
    \draw (2,12) node[below right]{$1$};
    \draw (4,14) node[right]{$1$};
    \draw (3,16) node[right]{$1$};
    \draw (1,16) node[left]{$1$};
    \draw (0,14) node[left]{$1$};
    \draw [dashed] (2,12)--(4,14);
    \draw [dashed] (4,14)--(3,16);
    \draw [dashed] (3,16)--(1,16);
    \draw [dashed] (1,16)--(0,14);
    \draw [dashed] (0,14)--(2,12);   
    \draw[dashed] (2,0)--(2,7);
    \draw[dashed] (4,2)--(4,9);
    \draw[dashed] (3,4)--(3,11);
    \draw[dashed] (1,4)--(1,11);
    \draw[dashed] (0,2)--(0,9);
    \draw [blue] (2,7)--(4,14); 
    \draw [blue] (4,9)--(2,12); 
    \draw [blue, dashed] (4,9)--(3,16); 
    \draw [blue, dashed] (3,11)--(4,14); 
    \draw [blue, dashed] (3,11)--(1,16); 
    \draw [blue, dashed] (1,11)--(3,14); 
    \draw [blue, dashed] (3,11)--(1,16); 
    \draw [blue, dashed] (1,11)--(3,14); 
    \draw [blue, dashed] (1,11)--(0,14); 
    \draw [blue, dashed] (0,9)--(1,16); 
    \draw [blue] (0,9)--(2,12); 
    \draw [blue] (2,7)--(0,14); 
  \end{tikzpicture}
  \caption{Bundle diagram of the KCBS experiment. The edges in blue colour correspond to possible correlations. Some lines have been removed for clarity reasons.}
  \label{fig:BundleKCBS}
\end{figure}
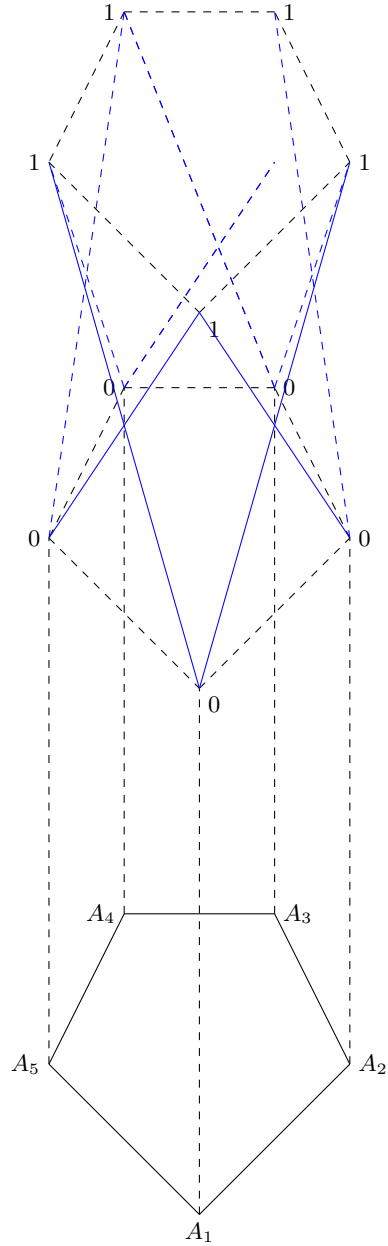

The last level of the hierarchy of contextuality is called \textit{strong contextuality}. It corresponds to a situation for which no global assignment consistent with the model is possible. The Peres-Mermin square and the KCBS scenario are strongly contextual. We give the possibility table of the KCBS scenario in Table~\ref{tab:SheafKCBS} and its bundle diagram in Figure~\ref{fig:BundleKCBS}. As we can see in Figure ~\ref{fig:BundleKCBS}, no univocal path can be drawn.

\begin{table}
\centering
\begin{tabular}{c c c c c}
            & 00  & 01  & 10 & 11 \\
\hline $A_1 A_2 $ & $\xmark$ & $\checkmark$ & $\checkmark$ & $\xmark$  \\
\hline $A_2 A_3 $ & $\xmark$ & $\checkmark$ & $\checkmark$ & $\xmark$  \\
\hline $A_3 A_4 $ & $\xmark$ & $\checkmark$ & $\checkmark$ & $\xmark$  \\
\hline $A_4 A_5 $ & $\xmark$ & $\checkmark$ & $\checkmark$ & $\xmark$  \\
\hline $A_5 A_1 $ & $\xmark$ & $\checkmark$ & $\checkmark$ & $\xmark$  \\
\end{tabular}
\caption{The possibility table of the KCBS experiment.}
\label{tab:SheafKCBS}
\end{table}

\subsection{A brief presentation of the mathematical frameworks of the model}
We connect the topological elements of the bundle diagrams to the mathematical notions that describe the experiments. We restrict ourselves to the mathematical elements that are of direct use for our study. 

\subsubsection{Correlation experiments in sheaf-theoretic approach}
\label{subsub:sheafcorrelationexperiments}
In the sheaf-theoretic approach, the elements that appear on the possibility tables are called \textit{measurement labels}, and are grouped into the set  $\mathcal{X}$. They are in finite numbers. They correspond to observables within quantum theory. The \textit{contexts} are noted $\mathcal{C}_i$ and they are subsets of $\mathcal{X}$. They are gathered in a \textit{maximal context} $\mathcal{M} = \{ \mathcal{C}_i \}_{i \in I}$. No maximal context can be included in another. Inside a context, all observables are compatible. A context is the set of all observables within the same line in the tables of possibility. All the observables have their outcomes in the same set $\mathcal{O} = \{o_0, o_1, o_2, o_3,...\}$, which is a fibre in sheaf theory. This leads to the following definition:
\begin{definition}{\textbf{Measurement scenario}}
We call \textit{measurement scenario} the knowledge of the triple  $\moy{\mathcal{X}, \mathcal{M} ,\mathcal{O}}$. It completely characterises an experiment.
\end{definition}
In the case of the Peres-Mermin square scenario, with notations presented in~\ref{sub:PM}, we have
\begin{equation*}
    \mathcal{X} = \{ A_{11}, A_{12}, A_{13}, A_{21}, A_{22}, A_{23}, A_{31}, A_{32}, A_{33} \}  
\end{equation*}
\begin{equation*}
    \mathcal{O} = \{-1,1\}
\end{equation*}
\begin{align*}
    \mathcal{M} = &\{ \{A_{11}, A_{12}, A_{13} \}, \{ A_{21}, A_{22}, A_{23}\} ,\{ A_{31}, A_{32}, A_{33} \}, \\ &\{A_{11}, A_{21}, A_{31} \},\{A_{12},A_{22},A_{32}\},\{A_{13},A_{23},A_{33}\} \}
\end{align*}

With regards to the previous approach, the major difference is the explicit use of the fibres of outcomes, whose topological properties are highlighted, whereas they were hidden in the graph approach. This property is crucial to highlight the different types of contextuality. 

\par The main objects that we study are called \textit{empirical models}. The easiest way to apprehend this notion is to  base ourselves on the definition~\ref{def:empiricalmodelprobability}, expressed in probabilistic terms. 
\begin{definition}{\textbf{Empirical model in probabilistic terms}}
An \textit{empirical model} is defined by a probability table associated to a measurement scenario, $\moy{\mathcal{X}, \mathcal{M} ,\mathcal{O}}$. The acquisition of data is led with repeated experiments for a fixed preparation. These dara are gathered in tables specifying probability distributions over the joint outcomes of sets of measurements that respect the \textit{compatibility principle of marginals}, according to 
\begin{equation}
e = \{ e_C \}_{C \in \mathcal{M}},
\end{equation} where $e_C$ is a probability distribution on a context $C$.
\label{def:empiricalmodelprobability}
\end{definition}

\subsubsection{Compatibility of marginals}
\label{subsub:Sheafcompatibility}
We can see that on any bundle diagram, every edge is connected to at least another one. This extension of edges corresponds to the compatibility principle, at a level of possibility \cite{Abramsky2017conf}. Indeed, for each measurement context $C$, there is a probability distribution $e_C$ on the set of functions that assigns an outcome in $O$ to each measurement in $C$. 
\begin{definition}{\textbf{Compatibility principle in the ABS framework}}
The requirement of compatibility of marginals is written mathematically
\begin{equation}
    \forall C, C' \ \in \mathcal{M}, \ e_ {C|C \cap C'} = e_ {C'|C \cap C'}
    \label{eq:Sheafcompatibility}
\end{equation}
\label{def:compatibility}
\end{definition}
where the notation $e_{C|U}$ with $U \subseteq C$  stands for the restriction of $e_C$ to $U$. The marginals of an observable must coincide when they emerge from two different contexts. This corresponds to a generalised no-signalling principle, in the sense that it matches this principle for space-like separated scenarii. However, this terminology can be misleading since compatibility and no-signalling are different notions. We will shed more light on this distinction in the last part of this work and come back to Eq~\eqref{eq:Sheafcompatibility} when we turn to quantifying contextuality in subsection~\ref{sub:ContextualFraction}.

\subsubsection{Hierarchy of contextuality}
\label{subsub:Sheafcontextualitydefinition}

The notions of this section can be defined with two aspects, a probabilistic aspect and a sheaf aspect. We will use one or the other, or both, according to the situation. It is our belief that the probabilistic aspects are in general more intuitive, and are suited to the definitions of compatibility. On the other hand, the sheaf structure is suited for the definition of the three levels of contextuality, particularly for the logical aspect~\cite{MansfieldConv}.

\paragraph{Probabilistic Aspect}
In terms of probability theory, an empirical model $e$ is \textit{contextual} if the family of probability distributions $e_C$ can not be obtained from the marginalisation of a probability distribution on global assignment of outcomes to all measurements. In mathematical terms, 
\begin{definition}{\textbf{Non-contextuality in probability terms}}
 $e$ is deemed \textit{non-contextual} if 
 \begin{equation}
     \exists \ d \in \mathcal{O}^{\mathcal{X}} \ | \ \forall \ C \ \in \mathcal{M}, d_C = e_C
 \end{equation}
where $\mathcal{O}^{\mathcal{X}}$ is the ensemble of applications from $\mathcal{X}$ to $\mathcal{O}^{\text{Card}(\mathcal{X})}$. If such a $d$ does not exist, $e$ is \textit{contextual}.
    \label{def:sheafcontextuality}
\end{definition}
In the case of a contextual model, the following classification can be established:
\begin{enumerate}
    \item \textit{Probabilistic contextuality} It coincides with definition~\ref{def:sheafcontextuality}. The families of distribution $e_C$ for all $C$ in $M$ can not be obtained as the marginals of a single distribution $d$ on $\mathcal{O}^{\mathcal{X}}$, such that for all $C$ in $M, d_C = e_C$. 
    \item \textit{Logical contextuality}
    \begin{itemize}
        \item We define new distributions $e_C$ with binary values, replacing all the strictly positive values of the former distribution by $1$ and keeping $0$ otherwise (this is the process by which we went from Table~\ref{tab:SheafProba} to Table~\ref{tab:SheafCHSH}). An empirical model is said to be logically contextual if it respects definition~\ref{def:sheafcontextuality} with this new distribution. 
        \item Mansfield suggests another approach~\cite{MansfieldConv}, more closely linked to the impossibility to close a global assignment. One could go one by one through the possible events, and in each case generate a new empirical model by post-selecting on the data to ignore any other events that are incompatible with it, and then checking probabilistic contextuality of the new post-selected models. Besides this way of proceeding links the contextuality discovered this way to the contextual fraction of the post-selected model (that we will only define later, in section~\ref{sub:ContextualFraction}).
    \end{itemize}
    \item \textit{Strong contextuality} An empirical model $e$ is said to be strongly contextual if there is no global assignment $g$ in $\mathcal{O}^{\mathcal{X}}$ such that for every context $C$ in the cover $M, e_C (g|C) > 0.$~\cite{Abramsky2017}.
\end{enumerate}

\paragraph{Sheaf aspect}
\textit{This part contains some technical explanations that will not be directly used in the rest of the document. The reader may skip it without consequences on the comprehension of what follows.}
\par There are scenarii for which contextuality arises only from the possibilistic tables and not from the probabilistic tables. To classify them, we need to present the sheaf aspect of contextuality. It gives all the mathematical explanations to the notions of \textit{global assignments coherent with the model} that corresponds to the univocal closed path in the bundle diagrams.
\par Consider a subset of the set of observables, $\mathcal{P} (\mathcal{X})$, and define a map $\mathcal{E}$ that goes from this set to a set of the set of outcomes $\mathcal{O}$, as shown in Equation~\ref{eq:sheaf}
\begin{align}
    \mathcal{E}:= \left\{
\begin{array}{lll}
\mathcal{P} (\mathcal{X}) & \rightarrow & \mathcal{O}^{\mathcal{X}} \\
                     \ U & \mapsto  & \mathcal{O}^U \end{array}
                     \right.
    \label{eq:sheaf}
\end{align}
where $\mathcal{O}^{\mathcal{X}}$ is a rather ill-defined ensemble, by which we mean the ensemble of the applications from $\mathcal{P} (\mathcal{X})$ into $\{ \mathcal{O}^n | n \leq \text{Card} (\mathcal{X}) \}$. This map is a sheaf, and is called the \textit{sheaf of events}. Each $s \in \mathcal{E} (U)$ is called a \textit{section}, so a section corresponds to the outcomes of some observables inside an experiment. When this section is the image by the sheaf $\mathcal{E}$ of all the observables inside $\mathcal{X}$, and not only of one of its subpart, it is called a \textit{global section}. This yields the important following definition:
\begin{definition}{\textbf{Consistency of a global section}}
 A global section $g$ is \textit{consistent} with an empirical model $e$ if 
 \begin{equation}
\forall \ C \ \in \mathcal{M}, \ e_C(g|_{C}) > 0 ,
 \end{equation} \textit{i.e.} when there are no negative probabilities.
\end{definition}

\par The notion of contextuality, that had arisen from logical paradoxes, and had been formalised as no-go theorems, is expressed in terms of structure. As such, an \textit{empirical model} can be defined with sheaf-theoretic terms.
\begin{definition}{\textbf{Empirical model in sheaf-theoretic terms}}
$e$ is an empirical model if it is a sub-presheaf of $\mathcal{E}$ for a given measurement scenario $\moy{\mathcal{X}, \mathcal{M} ,\mathcal{O}}$, for which 
\begin{itemize}
    \item $\forall \ C \in \mathcal{M}, e(C) \neq \emptyset$
    \item $\forall \ C \in \mathcal{M}$, if $U \subseteq U' \subseteq C$, $e(U') \rightarrow e(U)$ \text{is surjective}. 
    \item $e$ is compatible in the sense of definition~\ref{def:compatibility}.
\end{itemize}
\end{definition}
For $C \in \mathcal{M}$, and $e_c \in e(C)$, e is \textit{logically contextual} at $e_C$ if $e_C$ belongs to no compatible family, as defined in the previous section. The empirical model $e$ is \textit{logically contextual} if $e$ is logically contextual for some $e_C$, and \textit{strongly contextual} if it is logically contextual for all $e_C$. Another equivalent definition of strong contextuality is the fact that $e$ has no global coherent assignment. An empirical scenario which is contextual, but not logically contextual or strongly contextual, is \textit{probabilistically contextual}.

\subsection{New questions}
\label{sub:Sheafquestions}
Thanks to this new structure, several new questions have naturally arrived and have been addressed. As we saw, while entanglement is a property of a states, contextuality is characterised by a set of observables. An interesting question is consequently, given a state, what is the highest reachable level of contextuality as we range over all finite sets of measurements? A series of results have been obtained, among which the following theorems~\cite{Abramsky2017conf}: 
\begin{itemize}
\item \textbf{N-qubit pure states}: \textit{A n-qubit pure state admits measurements for which it is logically contextual if and only if it cannot be written as a product of one-qubit states and maximally entangled bipartite states.}
\item \textbf{Two-qubit}: \textit{No two-qubit state can achieve strong contextuality.}
\item \textbf{GHZ}: \textit{Only states in the  Greenberger, Horne, Shimony, and Zeilinger (GHSZ) SLOCC class (a stochastic extension to the well know  Local Operations and Classical Communication (LOCC) class) can achieve strong contextuality with any finite set of measurements. Moreover, these states must be of a constrained form and only equatorial measurements need to be considered.} ~\cite{Abramsky2017minimum}
\end{itemize}
Besides, the theory of cohomology has been successfully used to establish sufficient conditions of contextuality in a large number of scenarii. Finally, this formalism makes it possible to quantify the amount of contextuality of an empirical model \cite{Mansfield2017conf, Abramsky2017}, which will be crucial to experimentally test contextuality. We will thus present this quantification in details in section~\ref{NewBounds}. 

\subsection{An illustration of contextuality}
\begin{figure}
    \centering
    \includegraphics[width=1\linewidth]{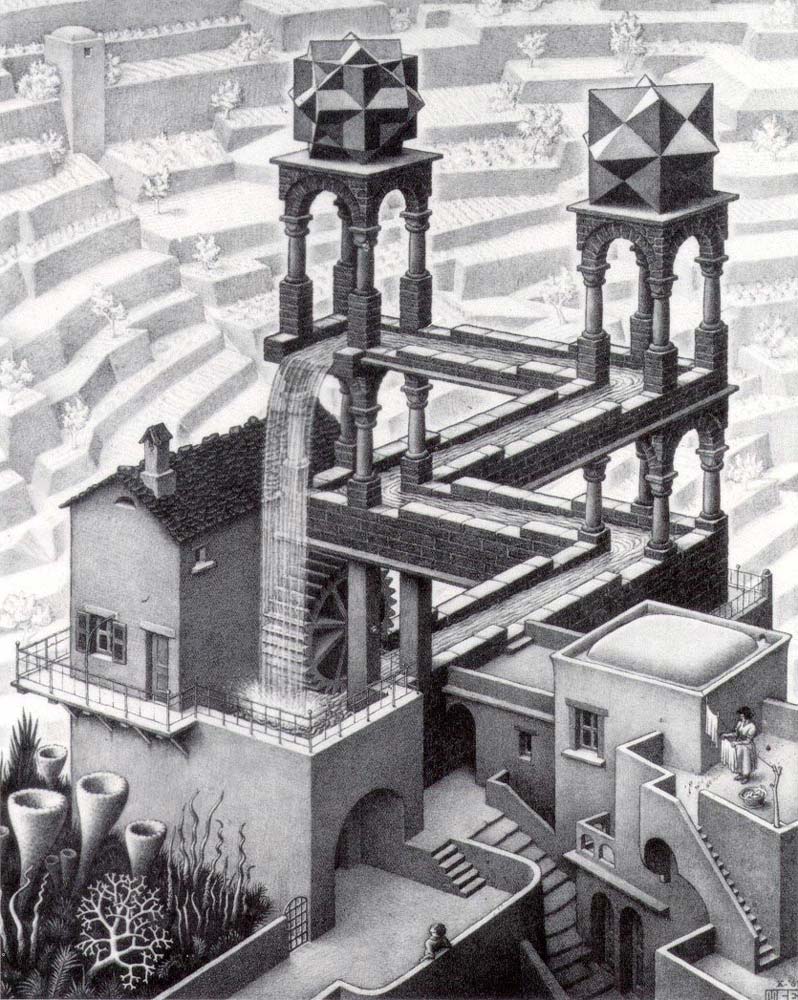}
    \caption{\textit{Waterfall}, a lithograph by M. C. Escher. The artist chose unrealistic proportions for the different part of his artwork. The result is an impossible figure, for which each elements are locally correct.}
    \label{fig:Escher}
\end{figure}

This structural approach of contextuality has highlighted the tension between local and global assignment. A well-known illustration of the impossibility to obtain a globally coherent model from locally well-defined events is given by the lithograph of Maurits Cornelis Escher called \textit{Waterfall}, which we give in Figure~\ref{fig:Escher}.


\section{A hypergrah approach to contextuality}
\label{ALFS}
The theory of hypergraph was the next step in the direction of establishing a general mathematical framework for contextuality, flexible enough to incorporate the previous approaches. It is claimed in the seminal paper of Ac\'in, Fritz, Leverrier and Sainz (AFLS)~\cite{Acin2015} that this approach comprises the graph-theoretic approach of section~\ref{CSW} and the sheaf-theoretic approach of section~\ref{Abramsky} as special cases.

\par The theory links with the works we mentioned before by studying test spaces as contextuality scenarii, and coordinating them with the Foulis-Randall product, a product able to cope with the empirical results of quantum physics while naturally taking into account the no-signalling principle~\cite{Sainz2017conf}. A hierarchy of probabilistic models is tested on this framework, among which non-contextual deterministic models and quantum models, and from them a hierarchy of bounds on inequalities is derived. The main advantage of AFLS theory is that it makes it possible to characterise accurately quantum correlations and quantum models. To do so, expectation values of quantum models established in different context are recorded and processed inside specific matrices, whose positivity properties establish a hierarchy of contextuality. Semidefinite programs (SDP) provide a computational, and sometimes analytical ressource to approximate the quantum set and establish whether or not a probabilistic model has a quantum realisation~\cite{SainzConv}. This is the transposition in the noncontextuality field of the hierarchy of semidefinite program for Bell scenarii, related non locality properties.

\par The main difference with the previous CSW model is that AFLS imposes that the sum of probability of outcomes is equal to $1$, not below $1$. This is done by taking into account the no-detection events. AFLS construction, just as CSW and Abramsky, apply both for contextuality and locality. Another significant achievement is the fact that AFLS yields the correct value for Tsirelson's bound~\cite{SainzConv}, contrary to CSW sometimes. Furthermore, AFLS construction is more suited to the study of composition rules for multipartite scenarii. 

\par Finally, this theory is different from the sheaf-theoretic approach in the sense that the former defines vertices of hypergraphs as measurement outcomes, while the latter defines them as observables, as seen previously. The different concepts of the two constructions are shown to correspond one with another. For instance, a bijection is defined between the empirical models on a given induced marginal scenario, from the sheaf theoretic approach, and probabilistic models from contextuality scenarii, from the hypergraph approach.


%


\section{An operational version of contextuality}
\label{SpekkensContextuality}
The last of the general theories that propose a general understanding of contextuality is that of Robbert Spekkens, and it was created in parallel with the others during the course of the 2000's and 2010's. Spekkens began by questioning the specificity of contextuality inside quantum mechanics. He first showed that most of the counter intuitive results of quantum mechanics could be obtained via classical statistical models. Only contextuality and non-locality could not be obtained this way. He then established a framework to study contextuality that goes beyond quantum mechanics, that reinterpretates previous noncontextuality inequalities and derives new ones, making use of an hypergraph-theoretic approach.

\subsection{A toy model} 
\label{sub:toymodel}
Physicists and philosophers have long discussed whether the representations of the world we created gave us access to the world in itself or only to the information we can acquire of it, a distinction famously established by Immanuel Kant under the terms of Noumenon and Phenomenon~\cite{Kant1787}. This distinction is translated into physical models by the concepts of ontic states and epistemic states. As clarified by Leifer, however, the question is not here about the nature of quantum states themselves, but rather their status \textit{within} the quantum theory itself~\cite{Leifer2014}. The nature of quantum states has been a long standing question for the community of physicists, because of the probabilistic nature of their measurement outcomes. Questions were also raised about the knowledge quantum states could gave of a system, a possible measurement of which being the fact that a state is a point (like in classical theory) or a cloud (like in quantum mechanics), inside a phase space representation~\cite{Spekkens2007}. In line with Fuchs interpretation of Quantum Mechanics~\cite{Fuchs}, Spekkens considered that quantum states were epistemic states of incomplete knowledge~\cite{Spekkens2007}. He noticed that in the smallest non-trivial set of ontic states, if he defined a smallest set of questions that could enable the observer to differentiate them, each question made him obtain as many information about the system as it would leave unknown. He turned it into a foundational principle~\cite{Spekkens2017conf}, and derived from it a hidden-variable theory that he put into correspondence with the Hilbert space framework of quantum mechanics. He was then able to show that phenomena such as entanglement, noncommutativity of measurements, teleportation, interference, the no-cloning and no-broadcasting theorems and unsharp measurements were present and easy to understand conceptually inside his model. The two notions that escaped it were locality and contextuality (see also~\cite{Jennings2015} for works in this direction).

\subsection{An operational theory of contextuality}
\label{sub:Spekkenscontext}
In order to study the particularity of contextuality, Spekkens presents an operational theory \cite{Spekkens2005}. Accordingly, it is not limited to quantum mechanics. Let us consider a black box with inputs, outputs and an agent that can process them. This can be the representation of any experimental situation. An operational theory specifies the probability of obtaining an output being given different possible inputs. It does not describe the agent. Spekkens explains the relevance of his operational approach by a reference to the \textit{Leibniz Principle}. Shortly, it states that if we are not able to distinguish between two objects, it is a fruitful approach for physicists to consider them equal~\cite{Spekkens2017conf, spekkens2019leibniz}.
\par Inside a laboratory, the different steps of an experiment can be conceptualised into three parts, preparation, transformation and measurement procedures. Thus, the operational theory specifies the probabilities $p(k|P,T,M)$ of obtaining outcome $k$, being given a preparation process $P$, a transformation $T$ and a measurement $M$. We can however subsume the transformation and the measurement $M$ inside only $M$, without lost of generality. When two preparations can not be distinguished, they form a class of equivalence, and can be represented by only one element. Indeed, 
\begin{equation}
 \left(  P \sim P' \right) \iff \left( p(k|P,M) = p(k|P',M) \ \forall \ (M,k) \right).
\end{equation}
The same can be applied to measurement processes, and yields
\begin{equation}
    \left( M \sim M' \right) \iff \left( p(k|P,M) = p(k|P',M) \ \forall \  (P,k) \right).
\end{equation}
The properties of an experiment that are not described by the specification of the class of equivalence of the preparation and measurement are called a context. In order to test contextuality, it is now necessary to present models of physical systems that are presumed to have properties regardless of any experiments or any agents, inside our theoretical framework. 
Such systems are called ontological, and they establish that the causal influence of the preparation on the measurement is mediated by the ontic state of the system, denoted by the hidden-variable $\lambda$, where $\lambda$ belongs to the ontic state space $\Lambda$. Inside an ontological model, preparations are assumed to prepare ontic states, and the same is true for measurements. 
Thus, given a preparation procedure $P$, we can define a probability density $\mu_P(\lambda)$ over the set of ontic states, and similarly, given a measurement $M$, we can define a probability of obtaining an outcome, $\xi_{M,k}(\lambda)$, called a response function. Both probability distributions are functions from $\Lambda$ to $[0,1]$, and they are such that
\begin{equation}
    \int \mu_P(\lambda) d \lambda = 1
\end{equation}
and, 
\begin{equation}
    \sum_k \xi_{M,k}(\lambda) = 1 \ \forall \ \lambda.
\end{equation}
Since the ontological model must reproduce the prediction of the operational theory, the probability of obtaining an outcome $k$, given a preparation $P$ and a measurement $M$ is
\begin{equation}
    p(k|P,M) = \int d \lambda \xi_{M,k}(\lambda) \mu_P(\lambda) \ \forall \ P, M.
\end{equation}

We can now define a noncontextual ontological model, one wherein the experimental results only depend on the equivalence classes of preparation and measurement, and not to its context. We introduce two essentiel notions for it.

\begin{definition}{\textbf{Preparation noncontextuality}}
A model is deemed \emph{preparation noncontextual} if it satisfies:
\begin{equation}
    P \sim Q \Rightarrow \mu_P(\lambda) = \mu_Q(\lambda)
\end{equation}
\label{def:pnc}
\end{definition}

\begin{definition}{\textbf{Measurement noncontextuality}}
A model is deemed \emph{measurement noncontextual} if it satisfies:
\begin{equation}
    M \sim N \Rightarrow \xi_{M,k}(\lambda) =\xi_{N,k}(\lambda).
    \label{eq:mnc}
\end{equation}
\label{def:mnc}
\end{definition}

\begin{definition}{\textbf{Universal non-contextuality}}
When an ontological model is both \emph{preparation} and \emph{measurement} noncontextual, it is called \emph{universally noncontextual}.
\end{definition}

Note that this analysis does not depend on any  particular  representation  of  preparations  and  measurements in an operational theory, nor on the particular probability rule associated with the theory; in this sense, it goes beyond quantum mechanics. Yet, we will apply it to QM, and show in the next sections how this formalism can shed a new light on our understanding of contextuality. 

\subsection{Application to quantum mechanics}
\label{sub:SpekkensAppliedQM}

If we apply this approach to quantum theory, the preparation $P$ is represented by a density matrix $\rho$ living in a Hilbert space $\mathcal{H}$. As a consequence, $\mu$ depends only on $\rho$, the measurements are represented  by  a  Positive  Operator Valued Measurement (POVM), $\{E_k\}$, such that 
\begin{equation}
    \xi_{M,k}(\lambda) = \xi_{\{E_k\},k}(\lambda)
\end{equation}
and the rule for assigning the joint probability is 
\begin{equation}
    p(k|M,P,\lambda, \lambda')  = Tr\left( E_k(M,\lambda') \rho(\lambda,P)\right)
\end{equation}
such that the $k^{th}$ element of $E_k$ is associated with the $k^{th}$ outcome. 

In the case of quantum theory, two measurement procedures  differ only by  context  if  and  only  if  they  yield the same statistics for  all  quantum  states.  In this case, they  are  represented  by  the same POVM \cite{Kunjwal2014}. Equivalently, two  measurement events differ only by context if and only if they are assigned the same probability by all quantum states. Since the measurements considered here are POVMs, the notion of commutation between observables is not anymore general enough, and it is replaced by a notion of joint measurement \cite{Yu2013, Kunjwal2014}. Note that a transformation procedure of the operational theory would be represented in QM by a completely-positive trace-preserving map.

\subsection{Reinterpretation of the Kochen-Specker results}
\label{sub:reinterpretation}
The framework of Spekkens gives rise to a reinterpretation of the previous proofs of contextuality. Once we have assumed \textit{Measurement Non-Contextuality} (MNC), (definition~\ref{def:mnc}), we can either consider that measurements events respond indeterministically to $\lambda$, in which case 
\begin{equation}
    \xi_{M,k}(\lambda) \in [0,1],
    \label{eq:oid}
\end{equation} or we can assume that the measurement events are deterministic functions of $\lambda$. This last assumption is called \textit{outcome determinism} (O.D.).
\begin{definition}{\textbf{Outcome determinism}}
A model is said to be \emph{outcome determinist} if it satisfies the following equation:
\begin{equation}
 \xi_{M,k}(\lambda) \in \{0,1\}.
    \label{eq:od}
\end{equation}
\label{def:od}
\end{definition}

 We clarify the previous notion of contextuality, notably used in the initial proofs and in section~\ref{CSW}, with the following definition~\cite{kunjwal2016thesis}:
\begin{definition}{\textbf{Kochen-Specker non-contextuality}}
A model is said to be Kochen-Specker (KS) \emph{non-contextual} if it assumes \emph{measurement noncontextuality}(definition~\ref{def:mnc}) and \emph{outcome determinism} (definition~\ref{def:od}). A system is KS-contextual if it rules out a KS non-contextual model.
\label{def:kscontextuality}
\end{definition}
The proofs of contextuality derived from Kochen-Specker initial proof can be logically formulated in this way:
\begin{equation}
    \textit{QM} \wedge \text{MNC} \wedge \text{OD} = \emptyset.
    \label{eql:KSO}
\end{equation}
Since the works of CSW and ALFS apply on every probabilistic models, thanks notably to the notion of operational equivalence between measurements, the status of their proofs can be logically formulated as
\begin{equation}
    \textit{PM} \wedge \text{MNC} \wedge \text{OD} = \emptyset,
    \label{eql:KSE}
\end{equation}
where $PM$ stands for probabilistic models. On the contrary, Spekkens is able to present proofs of contextuality without assuming \textit{outcome determinism}. He argues that these proofs are consequently the only ones able to actually dismiss noncontextuality experimentally. Indeed, the assumption of \textit{outcome determinism} can be understood as a restriction to idealised experiments. It was shown in \cite{Spekkens2014} that once \textit{preparation noncontextuality} is assumed, measurements  should  be assigned  outcomes  deterministically if and only if they are projective, \textit{i.e.} ideal. Kunjwal generalised the logic to all operational theories~\cite{Kunjwal2015}. It was also shown that KS noncontextual models are strictly comprised within preparation noncontextual models~\cite{Leifer2013} (in fact, there is even another ensemble that strictly encompasses KS and is strictly encompassed by preparation, called the maximally epistemic model). According to Spekkens the only proofs that can be applied to quantum mechanics with unsharp measurements, that is non projective POVMs, are the ones who do not make the O.D. assumption. We will see in the next and last part how this criticism can be taken into account. To do so, we shall need a last structural theory of contextuality, that we present in the next section.

\section{Contextuality-by-Default}
\label{Cbd}

Contextuality-by-Default (CbD) is the last major theoretical framework we present. It has been exposed in several articles~\cite{Dzhafarov2014, Dzhafarov2015, Dzhafarov2015ThreeTypes, Dzhafarov2016, Dzhafarov2016CBD2, Dzhafarov2020epistemic, Kujala2019, Dzhafarov2020cbdABS, Dzhafarov2020Cyclic}, a good possible sum-up being~\cite{Dzhafarov2015}. Like the operational theory of Spekkens, it is not restricted to quantum mechanics (in fact, it can be applied to others fields, such as human behaviour studies~\cite{Cervantes2019psycho}) but puts forward a bigger perspective on contextuality. It can be applied on any correlation experiment quite easily. We shortly sum it up in a not too technical way.

\subsection{Presentation of the theory}

The main philosophical argument of CbD consists in refusing to identify \textit{a priori} two measurement process in distinct contexts. In the case of Peres-Mermin square, it means that measuring $9$ different observables, each of them in two distinct contexts, leads to 
$18$ different mathematical objects, for which we will \textit{afterwards} study to what extent pairs of these $18$ objects can be statistically related, or connected to use CbD terminology. The origin of the name Contextuality-by-Default is that, by-default, two observables measured in two different contexts are deemed to be different in essence. 

\par The same observable, measured in two different contexts, do not have a “probabilistic relation", they are never “jointly distributed"~\cite{Dzhafarov2015}. However, a joint distribution, called a probabilistic coupling, can be computed, in order to match as much as possible these two distributions. In the special case where in the probabilistic coupling the random variables are always equal with probability one, the model is considered non-contextual. When such a coupling is not possible, the model is called contextual. This seemingly paradoxical approach yields a conceptually clearer and experimentally ready-to-use way to quantify the degree of contextuality of data, fit to take into account small seemingly paradoxical data results, that are believed to originate from measurement disturbance. 

\par Note that Winter had seemingly perceived this problem in~\cite{Winter2014}. He notes that we identify different outcomes from different measurements in two different ways in our theories. For the hidden-variable theory, two associated random variables take the same value, whereas in the quantum model the outcomes are deemed to represent the same effect in different measurements. CbD took the major conceptual step to base itself on disconnecting the identity of two random variables in different contexts. We now present more extensively the mathematical framework of CbD so that this document is self-contained, but encourage the interested reader to read it directly from the source~\cite{Kujala2015, Dzhafarov2015}. We stick to the notations and presentation of~\cite{Kujala2015, Dzhafarov2015}. We chose to indicate the meanings of CbD terminology in the quantum framework to help the reader. We will explicitly apply CbD theory on a given experiment in the last part of this work.

\subsection{Formal aspects and connection to other theories}

For the purpose of a correlation experiment, a finite set of distinct physical properties $Q = \{q_1,...q_n \}$ are measured in subsets of $Q$. These subsets, called \textit{contexts} and denoted $c_1,...c_m$, regroup physical properties that can be measured together. Indeed, in quantum theory, \textit{physical properties} are observables, and the impossibility to be measured together is incompatibility. We denote $C$ the set of all contexts and $C_q$ the set of all contexts containing a given property $q$.  The measurement results of a property $q$ in a context $c$ are a random variable $R_q^c$. This variable is the response function in Spekken's framework. The already mentioned crucial assumption of CbD is founded on the fact that two random variables in different contexts, $R_q^c$ and $R_{q'}^{c'}$, with $c \neq c'$, are never jointly distributed. The next definition is crucial:
 \begin{definition}{\textbf{Connection}}
A \textit{connection} is a set of random variables representing the same property $q$ in different contexts: $\{R_q^c : c \in C_q \}$. 
 \end{definition}

In CSW, connections would regroup the same observables. A system is the disjoint union of all random variables taken in each of their context. 

Now, we try to see to what extent it is possible to “overlay" two distinct random variables within a connection thanks to only one random variable. 

\begin{definition}{\textbf{Coupling of a connection}}
With $q$ a fixed property, a coupling of a connection $\{R_q^c : c \in C_q \}$ means that there exists a jointly distributed $\{S_q^c : c \in C_q \}$ such that 
\begin{equation}
    \forall c \in C_q,\  S_q^c \sim R_q^c,
    \label{eq:connecoupling}
\end{equation}
\end{definition}
where $\sim$ means “has the same joint distribution". 
It is always possible to define a joint distribution on a system but we will be looking for the more appropriate. Let us extend this concept to the whole system:
\begin{definition}{\textbf{Coupling}}
A coupling is a jointly distributed set of random variables $S = \{S_q^c : q \in c \in C \}$ such that, 
\begin{equation}
   \forall c \in C, \ \{S_q^c : q \in c\} \sim \{R_q^c : q \in c  \}
    \label{eq:coupling}
\end{equation}

\end{definition}
and contextuality is defined thanks to:
\begin{definition}{\textbf{Non-contextuality in CbD (without signalling)}}
If there is one coupling $S$ for which for all property $q$, the random variable obtained by marginalisation of $S$ are equal within all different contexts with probability $1$, there is a noncontextual description of the system. 
\end{definition}
The decisive computation to make is consequently, for every coupling $S$ of the random variables $R_q^c$, for every property $q$,
\begin{equation}
    \text{Pr}[S_q^{c_{q1}} = ... = S_q^{c_{qn_q}}], \{ c_{q1},...,  c_{qn_q}\}.
    \label{eq:noncontextualitycouping}
\end{equation}

If there is no such coupling, the system is \textit{contextual}. This is similar to the ABS theory, where a system is contextual when a global assignment can be made that yields all elements in contexts by marginalisation.

\subsection{Taking NS into accout and quantifying contextuality}

 One may have noticed that an hypothesis has been made, which is called \textit{consistent connectedness} :
 \begin{definition}{\textbf{Consistent connectedness}}
 A system is called consistently connected when all random variables within each connection are identically distributed.  This property is also called the \textit{Gleason property}.
 \end{definition}
 This is a requirement in ABS, and the reason why the notions of contextuality have been so far equivalent~\cite{Dzhafarov2020cbdABS}. A major result of the CbD framework is to be able to bypass this hypothesis. If the system is \textit{inconsistently connected}, then there would necessary exist some property $q$ and some context $c,c'$, for which $\text{Pr}[S_q^c = S_q^{c'}]$ can not be equal to $1$, because the distribution of $R_q^c $ and $R_q^{c'}$ are not even the same. Shall such systems be necessarily deemed \textit{contextual}? No, because then \textit{contextuality} would be conditioned on obtaining perfectly similarly distributed random variables within connections, which is empirically impossible. That is why we now extend the concept of \textit{contextuality}.

\begin{definition}{\textbf{Maximal non-contextual description}}
A system is said to have a \textit{maximal noncontextual description} if there is a coupling $S$ of $R_q^c$ such that for any property $q$, the random variables $\{ S_q^c : c \in C_q\}$ are equal to each other with the maximum probability allowed by the individual distributions of $R_q^c$.
\end{definition}
Let us give a few more details to better understand this extension. 
\begin{definition}{\textbf{Maximal coupling}}
A coupling of a connection is said to be maximal when the probability that all variables in the joint distribution $S_q$ considered with regard to the elements of the connection is the maximal possible. 
\end{definition}
To say it more clearly, it is the best possible joint distribution one can impose on a connection. Mathematically speaking, in the case of binary measurements $\pm 1$, the maximal coupling $T_q$ is the one for which
\begin{equation}
\text{eq}(T_q) = \text{Pr}[T_q^{c}=1 \ : \ c \in \mathcal{C}]+\text{Pr}[T_q^{c}=-1 \ : \ c \in \mathcal{C}]
\end{equation}
is the maximum. This quantity is always well defined. The hypothesis of \textit{consistent connectedness} imposed that it was equal to $1$. In the general case that we now consider, it can take any value in $[0,1]$. Once again, we now move to the global system with the following definition:

\begin{definition}{\textbf{Maximal connectedness}}
A coupling $S$ is said to be maximally connected if its restriction by marginalisation to any connection is maximal.
\end{definition}
We can now define contextuality.
\begin{definition}{\textbf{Contextuality in CbD}}
A system is said to be contextual if no coupling of this system is maximally connected.
\end{definition}

The empirical correlations results are compared to the probability distributions of observables recorded separately, and a quantitative criterion is given to evaluate this new concept of \textit{noncontextuality}. This new description is fit to handle the small deviations to noncontextuality recorded in experiments and paves the way to an experimentally robust quantification of contextuality. The idea is to take a difference between sum of local assignments and global assignment. We take a system $\mathfrak{R}$ and define:
\begin{equation}
\text{CNTX}(\mathfrak{R}) = \sum_{q \ \in \ Q} \max_{T_q} \text{eq}(T_q) - \max_{S} \sum_{q \ \in \ Q} \text{eq}(S_q)
\end{equation}
where for the first term we optimise with respect to possible couplings for every connection, and for the second for a global coupling to the whole system.

\subsection{CbD 2.0}
Very recently, the authors of CbD have derived a new version of Cbd, called CbD 2.0~\cite{Dzhafarov2016CBD2, Dzhafarov2017multi}. It is a  slightly different generalisation of contextuality. Let us call subsystem a system for which some random variables have been dropped.In consistently-connected systems, CbD has the property that any subsystem of a noncontextual system is noncontextual, but it is not the case for inconsistently connected-systems. CbD 2.0 chooses to extend this property. To do so, it replaces the notion of \textit{maximal coupling} with that of \textit{multimaximal coupling}: instead of looking for the maximal coupling for the connections, it looks for the maximal coupling for any subset of the connections. For technical reasons, it is then necessary to consider categorical random variables (random variables with a finite number of possible outcomes) and to dichotomize them. Another extended version of contextuality which chooses not to extend this property (because of experimental considerations) has been provided in~\cite{Amaral2019CbD}. Its results immediately apply to any experimental scenario (not only cyclic systems) and its derivations should be run faster on a computer.

\par In everyday words, CbD is interpreted as the capacity that the contexts have to force the variables to be more dissimilar than they are when taken in isolation~\cite{Dzhafarov2020QCQMB}. CbD and CbD 2.0 coincide in a large class of systems, notably the consistently connected ones and the cyclic systems~\cite{Dzhafarov2020Cyclic}. The case of the Peres-Mermin square that we shall consider in detail below is not directly among them, but since its connections contain only two random variables, the definitions still coincide.

\clearpage
\begin{widetext}

\section{Connecting the theories}
\label{comparisons}

\subsection{Definition of contextuality}
Contextuality is defined differently according to the approach retained. In the standard quantum approach,it is the impossibility to reproduce a set of statistics emerging from the quantum framework thanks to a hidden-variable model which assumes outcome determinism and measurement non-contextuality. This has been extended in the graph-theoretic approach into a systematic approach of the difference between noncontextual models and quantum (and even more general theories) models. In the operational framework, it is the non-existence of a hidden-variable model, called ontological model, which respects preparation non-contextuality and measurement non-contextuality, to reproduce some experimental statistics, no longer restricted to quantum mechanics. If the sheaf-theoretic approach sees contextuality as the non-existence of a joint probability distribution over the outcomes of a set of measurements, CbD disjoints the entities which represent the same measurement in two different contexts. It then quantifies to what extent it is possible to impose a probabilistic coupling over these results that matches and sometimes generalises previously considered definitions of contextuality. 

\subsection{Systematic comparison of the notions}
In order to compare the different theories, we base ourselves on three articles, which respectively highlight the connections between the:
\begin{itemize}
    \item Graph theoretic and sheaf-theoretic approach~\cite{Silva2017}
    \item Sheaf-theoretic and Operational approach~\cite{Wester2018}
    \item Sheaf-theoretic and CbD approach~\cite{Dzhafarov2020cbdABS}
\end{itemize}
The following table is conceived as a way to depict the correspondence between the different approaches in a loose way. Indeed, some notions put in correspondence with others are \textit{more general}, and some are even structurally different. This table is only conceived as a guidance for the reader to connect more easily the theories, in no case should it be perceived as any kind of isomorphism between them. To see in details to what extent such isomorphisms hold and entail a possible unified version, we refer the reader to an in-depth reading of the articles quoted above. Finally, let us note that a contribution has also clarified the links between the graph-theoretic approach and the CbD theory~\cite{Amaral2018}.

\begin{table}[h]
\begin{tabularx}{\textwidth} { 
  | >{\centering\arraybackslash}X 
  | >{\centering\arraybackslash}X 
  | >{\centering\arraybackslash}X 
  | >{\centering\arraybackslash}X 
  | >{\centering\arraybackslash}X 
  | >{\centering\arraybackslash}X |}
 \hline
 \textbf{Quantum} & \textbf{Graph} & \textbf{Sheaf} & \textbf{Operational} &  \textbf{CbD}\\
 \hline
Reference & Systematic derivation of inequalities & Hierarchy  & Black-box model  & Probabilistic Coupling \\
\hline 
Density matrices &  & State $\sigma$ & Equivalent classes of preparation & Random variable \\
\hline
POVM &  & Measurement labels & Equivalent classes of measurement & Coupling on random variables from the same connection \\
\hline
PVM & Vertex &  & Equivalent classes of measurement with Outcome Determinism &  \\
\hline
No-signalling &  & all states respect : $\sigma_{C|C \cap C'} = \sigma_{C'|C \cap C'} $ & Outcome Determinism &  Consistent-connectedness \\
\hline
Signalling &   & $\exists (\sigma, C, C') | \sigma_{C|C \cap C'} \neq \sigma_{C'|C \cap C'} $ & Outcome Undeterminism & Inconsistent-connectedness \\
\hline
Ensemble of joint-wise but not triple-wise observables & Exclusivity graph & Measurement scenario & Preparation, Transformation and Measurement scenario & Finite system of dichotomous random variables \\
\hline
Context (set of observables, originally commuting) & Adjacent vertices & Maximal set of compatible measurements & Jointly measurable POVM & Jointly distributed random variables \\
\hline
\end{tabularx}
     \label{tab:comparison}
\caption{Connections between the structural theories of contextuality - The second line represents the “main idea" in our view. Note that the graph model is used for both the quantum and the classical model, which makes this column rather ambiguous. This table is my opinion, any comment or clarification shall be welcome.}
\end{table}

\clearpage
\end{widetext}


\begin{center}
\textbf{\textsc{\Large The quest to derive experimentally robust proofs of contextuality}}
\end{center}

Once proofs of contextuality for Quantum Mechanics were obtained, physicists tried to turn them into experimental tests. That is one of the reasons why noncontextual inequalities, such as the ones already seen, were derived. However, the transition from mathematical proofs suited to idealised situations to realistic ones, adapted to noisy environment, raised several conceptual issues. We try in what follows to explain how robust inequalities can be derived, and to clarify their scope. The heart of our work consists in circumventing the criticism of Spekkens within the framework of Quantum Information.

\par We carefully analyse the problems that arise when the inequalities are adapted to noisy environment in section~\ref{Analysis}. We then explain how new bounds for noncontextual inequalities can be derived, so as to be robust to noise, in section~\ref{NewBounds}. Lastly, we explain how we can encompass the experimental results within quantum theory in section~\ref{NoiseforPM}. We then conclude in section~\ref{FinalProtocol} on the scope of these experiments, with notably a quantum foundational perspective. 
\par We chose to stick to a particular non contextuality inequality, the one based on the Peres-Mermin square, already presented in section~\ref{sub:PM}. We deem this inequality as particularly in line with our wishes to establish robust and general inequalities, thanks to its state-independent aspects, although this property slightly vanishes when we take noise into account, as we will see. However, the ideas we present and the calculations we make apply in the same way to the other inequalities that can be cast in the form presented in section~\ref{CSW}.


\section{Analysis of the loopholes in the experimental determination of contextuality}
\label{Analysis}
The ability of KS-contextuality inequalities to produce experiments that show that nature is contextual has been attacked by two major controversies. We analyse them successively in this part, beginning by the Finite-Precision controversy in section~\ref{sub:MKC}, before explaining the scope of the hidden connection between contextuality and outcome determinism criticizd by Spekkens, in section~\ref{sub:consequencesodum}. Based on this study, we justify our conception of contextuality as a quantifiable continuous resource~\ref{sub:LogvsEmp}, and distinguish two types of noise able to destroy this resource (section~\ref{sub:noisebetweenwithin}). The inequalities presented so far were established on the basis of idealised measurements. If these logical proofs could not be questioned and have a scope of their own~\cite{cabello1999comment}, a naive translation to experiments poses danger.

\subsection{The finite precision controversy}
The first issue arises from  the impossibility to reach a perfect projective measurement in experiments, or, even worst, from the impossibility to distinguish in a protocol a projector from an arbitrary close POVM~\cite{Winter2014}. 

\subsubsection{The Meyer Kent Clifton (MKC) model: introduction and reactions}
\label{sub:MKC}

Meyer showed how to approximate, in an arbitrarily close fashion, the set of uncolourable vectors used in the Kochen-Specker demonstration (see section~\ref{sub:KS} and~\cite{KochenSpecker1967, Szangolies2015}) by a 2-colourable sets of rational vectors~\cite{Meyer1999} . It was then generalised to arbitrary Hilbert spaces~\cite{Kent1999} and finally, Clifton derived a model that reproduced QM prediction with noncontextual hidden-variables~\cite{Clifton2000}. More precisely, the Meyer Kent Clifton (MKC) model showed that in every Hilbert space of a quantum system in dimension $d$, there exists a dense set of measurements that has a non-contextual hidden-variable model reproducing the correct statistic of any given state~\cite{Winter2014} measured by any POVM. The dramatic consequence is clear: no experimentalist can distinguish the observation that he or she makes, and that they deem to be quantum POVMs, from the ones they would have with non-contextual models, due to this finite precision. The attitudes towards this controversy were diverse, ranging from denying the plausibility of MKC sets to deriving new proofs not subject to the criticism (see ~\cite{Szangolies2015, Cabello2017conf} for reviews) and they engaged Meyer Kent and Clitfon in a long debate~\cite{Barrett2003}. In it, the reader can find a discussion on the relevance of the MKC model. The authors argue in favor of the physical acceptability of their counter-model by explaining that it can reproduce the QM results with only a finite number of discontinuities and that their model is classical. Indeed, Appleby had rejected as their hidden variable model as being non obtainable empirically~\cite{Appleby2005}. Barrett and Kent also dispute the call that, from the non-locality of the MCK model, one can retrieve a notion of contextuality inside the hidden variables model~\cite{Appleby2002}.

\subsubsection{Analysis of the precision loophole}
First, it should be noted that the link between the empirical outcomes and the theoretical observables that give birth to them can not be empirically proven. It is a principle of quantum mechanics. In~\cite{Winter2014}, Winter proposed a kind of continuity argument to tackle the MKC issue. He proposed to estimate (via a tomography) how close the quantum measurement is to an ideal projector, and to link it with how close the statistical outcome of a measurement is in two different contexts~\cite{Budroni2021}. We shall detail it further below (section~\ref{sub:Winter}), but it is clear that such a linking does not seem to be experimentally accessible, since the two objects come from different theoretical frameworks.
\par A very careful analysis of the loophole has been presented by Hermens~\cite{Hermens2014} whose main lines we shall here present. We invite the reader to consult the original article. Hermens translates the Kochen-Specker theorem in the following logical form:
\begin{equation}
    \text{QM} \wedge \text{Re} \wedge \text{FM} \wedge \text{CP} = \emptyset
    \label{eq:logicalks}
\end{equation}
where QM stands for Quantum Mechanics, Re for Realism (every observable possesses a certain value at all times), FM for Faithful Measurement (a measurement reveals the value possessed by that observable) and CP for Correspondence Principle~\cite{Hermens2014}.

\begin{definition}{\textbf{Correspondence Principle}}
There is a bijective correspondence between observables and self-adjoint operators
\end{definition}Moreover, the Correspondence Principle is divided into two distinct notions~\cite{Hermens2014}:
\begin{definition}{\textbf{Non-contextuality (NC)} -Hermens}: Every observable is uniquely defined by a self-adjoint operator
\label{def:NCHermens}
\end{definition}
and
\begin{definition}{\textbf{Identification Principle (IP)}}
Every self-adjoint operator represents an observable.
\end{definition}

MKC shows that it is in fact possible to construct an empirical noncontextual model that assumes Re and FM, and only rejects IP. Hermens shows a clear mathematical proof and proposes a time evolution to his model. He addresses the classical reproach to the MKC models, namely that the slightest measurement difference yields a totally different outcome, by imposing a quantum mechanical-like projection postulates for his hidden variables. It is not however clear to the author of this review if this construction, which certainly may work, has been proven to work. In any case, this work clarifies the unspecified hypothesis underlying our comprehension of contextuality.

\subsection{The Outcome Determinism for Unsharp Measurements (ODUM) controversy}
\label{sub:consequencesodum}

In section~\ref{sub:reinterpretation}, it was shown that Spekkens highlighted the importance of \textit{outcome determinism} in the derivation of so-called proofs of KS-contextuality according to the terminology of definition~\ref{def:kscontextuality}. Indeed, the bounds on the inequalities obtained in protocols such as~\cite{Cabello2008} are based on O.D. In the framework of this hypothesis, it is not possible that the value of an observable changes, between two different contexts, inside the same set of an experiment. However, these kind of deviations are experimentally observed~\cite{kirchmair2009}. They can be due to noise, defects of the recording apparatus, to the convergence in our hypothesis of the Law of large numbers~\cite{MansfieldConv}, etc.

\subsubsection{Refutation of the ability to reflect experimental scenarii with ideal measurements}

On the basis of these considerations, the ability of KS noncontextual inequalities to verify experimentally the contextuality of a model is allegedly dismissed. Indeed, in the Appendix C of~\cite{Krishna2017}, it is argued that the slightest variation in the value of one of the $9$ $A_{ij}$ observables that consitute the Peres-Mermin square can lead to a full violation of the noncontextual bound. In this case, any noncontextual model can reach the quantum bound of $6$. Let us explain this crucial point more explicitly. Remember the Peres-Mermin combinations of operators
\begin{align}
\moy{S} &= \moy{A_{11}A_{12}A_{13}} + \moy{A_{21}A_{22}A_{23}} +  \moy{A_{31}A_{32}A_{33}} \nonumber \\  &+\moy{A_{11}A_{21}A_{31}} +  \moy{A_{12}A_{22}A_{32}} - \moy{A_{13}A_{23}A_{33}}
\end{align}
If we assume Measurement Non-Contextuality, this turns into:
\begin{align}
\moy{S} &= \moy{A_{11}}\moy{A_{12}}\moy{A_{13}} + \moy{A_{21}}\moy{A_{22}}\moy{A_{23}} \nonumber \\ &+  \moy{A_{31}}\moy{A_{32}}\moy{A_{33}}  +\moy{A_{11}}\moy{A_{21}}\moy{A_{31}} \nonumber \\ &+  \moy{A_{12}}\moy{A_{22}}\moy{A_{32}} - \moy{A_{13}}\moy{A_{23}}\moy{A_{33}}
\end{align}
Now, the assumption of Outcome Determinism implies that if an observable is determined by a hidden variable, $A_{ij} = A_{ij}(\lambda)$, it can take the value $\pm 1$ but can not change according to the context. It is this hypothesis that restricts the number of possible combinations of $\moy{S}$ to $2^9$. If this hypothesis is not respected, as it is indeed the case in experiments, the bound $\moy{S} \leq 4$ does not hold anymore. In fact, any flip can lead to a maximum violation of the non-contextual bound $\moy{S}$. Indeed, let us consider the case where all $A_{ij}$ are positive: $\moy{S} =4$. Now let us suppose that $A_{13}$ is positive in the first context and negative in the second one. In this case, $\moy{S} =6$.

\par \cite{Krishna2017} also argues that one can not control the effect of these mistakes through statistical arguments, however rare it can be. While it is true within the framework he exposes, we exposes how the other frameworks can guarantee this control. Finally, the last criticism that was made was the fact that inside the classic CSW approach, no matter how noisy the states are considered, a violation could still be obtained. In contrast, a version of Peres-Mermin square based on Spekkens theory was proposed, and it was shown that in that case \textit{universal contextuality} could be proven, even in experiments where noise is taken into consideration~\cite{Krishna2017, Mazurek2017, Mazurek2016}.

\subsubsection{Refutation of the possibility to use POVMs and still assume OD}
\label{subsub:odum}

In this version, POVMs are used and justified by the fact that the operational framework directly takes into account outcome indeterminism. In contrast, several propositions were made to salvage the possibility of implementing POVMs while keeping the outcome determinism hypothesis. Most of them relied on Naimark extension - the fact than any POVM can be written as a PVM in a space of higher dimension - and Fine's result about Bell models - that indeterministic models yield the same results that probabilistic results in Bell scenarii~\cite{Fine1982}. They were refuted in~\cite{Spekkens2014}, drawing by the same process a distinction between Bell models and KS-noncontextuality models. 

\par Indeed, Outcome Determinism for Unsharp Measurements (ODUM) to use Spekkens terminology, relies on the belief that the distinction between an indeterministic model and a deterministic one is not crucial. Why is it so? Everyone agrees that outcome determinism is compatible with sharp measurements, a.k.a. PVMs. Now, consider a POVM. We can always represent this POVM by a PVM on a composite of the system and an ancilla (this is Naimark theorem). However, ODUM makes the false hypothesis that the response function associated with the PVM is unique. Since two PVMs can reduce to the same POVM via the tracing out of the ancilla, it means that the two PVMs are generally distinguishable - and are represented by different response functions. As such, they are governed by different hidden variables in a noncontextual scenario (the assumption of measurement non-contextuality collapses), and so no classical bound can be derived. This is just a sum-up, in non-technical language, of the proof derived by Spekkens in~\cite{Spekkens2014}. 

\par It was however argued that instead of taking the Spekkens attitude of considering noise as fundamental, it could be treated just as a small deviation from the theory~\cite{Cabello2017conf}, and correct the experimental mistakes with additional terms. Through these considerations, the notion of contextuality was gradually moving from a binary principle to a quantitative grading \cite{Abramsky2017}.

 \subsection{Logical proofs and empirical proofs}
 \label{sub:LogvsEmp}
Inspired by the MKC proposition, the operational framework defined contextuality beyond quantum mechanics. This theory was immediately compatible with experimental procedures, and able to produce “Nature vs Noncontextuality" experiments, instead of “Quantum Mechanics vs Noncontextuality" (which somehow did not suceed). The  Kochen \& Specker theorem had other implications, namely that the Quantum Mechanics formalism possessed this particularity. Spekkens successfully showed that the theoretical proofs of contextuality based on ideal measurements were improper for realistic setups, and more importantly that outcome determinism was incompatible with POVMs. This left as a possible solution to implement outcome undeterminism within the quantum framework. On this basis, we organise our inquiry on contextuality in two steps:
\begin{itemize}
\item Derive bounds for hidden-variable noncontextual models based on empirical data (section~\ref{NewBounds}).
\item Recover empirical results within quantum theory by taking noise into account (section~\ref{NoiseforPM}).
\end{itemize}
   
Doing so, we notably seek to evaluate the amount of contextual behavior in an experiment (with respect to the most general noncontextual classical framework). We also try to recover the experimental results within a quantum framework, and will discuss possible interpretations in section~\ref{FinalProtocol}.

\subsection{Within or between the contexts: a classification of noise}
\label{sub:noisebetweenwithin}

The noncontextuality inequalities are maximally violated in ideal systems based on PVMs and perfectly compatible observables within a context. We believe that the CbD approach clairifies the two sources of noise that can affect our experimental proofs. There is noise \textit{within} a context, because observables are never perfectly compatible, and noise \textit{between} the contexts, because there is always some source of signalling. This last one appears explicitly in the CbD theory when a connection is established between two observables in different contexts. Compatibility issues appears even when we are restricted to consistently-connected systems (see~\ref{tab:comparison}). We hope that disentangling these two sources of noise helps to clarify the experimental issues in the following part.


\section{New bounds for hidden-variable models based on empirical data}
\label{NewBounds}
In this section, we address the question of deriving new bounds for hidden-variable models that do not contradict empirical data. We focus on the Peres-Mermin square experiment, and present the solutions to the two previously derived controversies within four different approaches. 

\par In~\ref{sub:Winter}, we present the solution of Winter to the MKC controversy, and try to see to what extent it can apply also to the problem raised by Spekkens about the CSW theory. We then briefly expose how the Peres-Mermin square is treated within the operational framework in~\ref{sub:PMoperational} and discuss its robustness to MKC. We analyse the limitations of the Contextual Fraction, a quantifier of contextuality that comes from the Sheaf Theoretic approach, to tackle the no-signalling events~\ref{sub:ContextualFraction}. Doing so, we clarify the confusion between compatibility and signalling mentioned in section~\ref{subsub:Sheafcompatibility}. Finally, we apply the CbD approach in the case of the PM square~\ref{sub:CbDPM}, following the steps used for similar cases in~\cite{Dzhafarov2015}. We quantify the amount of contextuality of the Kirchmair \textit{et.al} experiment~\cite{kirchmair2009} within this framework. We conclude in section~\ref{sub:CclNB} on the relevance of these propositions to adequately answer the two previously considered controversies, and to disentangle the sources of noise within contexts (compatibility) and between the contexts (no-signalling).

\subsection{The Peres-Mermin square in a CSW approach}
\label{sub:Winter}

\subsubsection{Motivations and scope}
In \cite{Winter2014} a framework was established to compute new bounds for noncontextual hidden-variable models that closes the MKC loophole for the CSW approach. It actually answers the Spekkens controversy. The model developed below assumes outcome undeterminism, and shows how this deviation from the ideal scenario changes the classical bound.

\subsubsection{A formalism that incorporates statistical deviations}

We present the formalism used in~\cite{Winter2014}. We begin by the following definition:
\begin{theorem}{($\epsilon$-ONC) model}
An $\epsilon$-ontologically faithful non-contextual ($\epsilon$-ONC) model for a hypergraph $\Gamma$ of contexts $C \in V$ consists of a family of random variables $X_i^C \in \{0,1\}, i \in C \in \Gamma$, such that 
\begin{align*}
    &\forall \ C \in \Gamma, \sum_{i \in C} X_i^C \leq 1  \\
    &\forall \ C,C' \in \Gamma \ \forall i \in C \cap C', \text{Pr}\{X_i^C \neq X_i^{C'} \} \leq \epsilon
\end{align*}
\end{theorem}
The probability that a measurement outcome takes two different values in two different contexts is then upper-bounded. 
Let us recall a general expression of a noncontextual inequality, adapted to our formalism. We can write it
\begin{equation}
\sum_i w_i X_i \leq \alpha
\label{def:NCIW}
\end{equation}
where $w_i$ are the weights of the outcomes $X_i$ and $\alpha$ is the maximum of the left part of the inequation, over all noncontextual hidden variable models.

\subsubsection{Inequalities adapted to experimental errors}

Now, if we want to adapt this inequality to the small deviation we quantify above, we define a noncontextual hidden variable model as
\begin{equation}
    Y_i := \prod_{i \in C \in \Gamma} X_i^C
\end{equation}
Note that this product is made for a given outcome $i$ over all possible contexts.
With $k_i$ the number of times an outcome i occurs in some context $C \in \Gamma $, 
\begin{equation}
    \text{Pr}\{\exists \ i \in C s.t. X_i^C \neq Y_i\} \leq (k_i - 1) \epsilon
\end{equation}

With that, the new hidden variable noncontextual inequality will be replaced by 
\begin{equation}
    \sum_i \lambda_i Y_i \leq \alpha + \epsilon \sum_i \lambda_i (k_i - 1)
    \label{eq:NCINewBound}
\end{equation}
and the quantity on the right part of inequation \ref{eq:NCINewBound} is the new bound to violate for quantum theory results.
From this formalism, it is possible to establish a new bound for the Peres-Mermin inequality of inequation \ref{eq:PM}, that will be related to the experimental data. 

\par In this formalism, it is possible to propose an adapted inequality for the Peres-Mermin square. According to~\cite{Winter2014}, we can reconsider inequality~\eqref{eq:NCINewBound}, and, knowing that the quantum bound is $6$, we obtain:
\begin{equation}
    5 + \epsilon \times 72 \leq 6.
    \label{eq:PMWinter}
\end{equation}
With an appropriate $\epsilon$, MKC models could be excluded. However, as already mentioned, it is not experimentally accessible.

\subsection{Peres-Mermin square from an operational approach}
\label{sub:PMoperational}
We now present the operational version of the PM square. We present it in the next subpart, then discuss with the help of Hermens its capacity to close the finite precision loophole.

\subsubsection{Presentation}
This subpart is extracted from~\cite{Krishna2017}. This article provides a general protocol to turn KS-contextual inequalities into universal inequalities. Here we are only interested in the PM case. First, the article translate the PM inequality into sharp PM universal non-contextuality inequality, then a depolarizing noise map is introduced:
\begin{equation}
    \mathcal{D} = r \mathds{1} + (1-r) \frac{1}{4} \mathds{1} \Tr
    \label{eq:depolPM}
\end{equation}
where $r \in [0,1]$ quantifies the amount of noise in the system. It then turns to presenting inequalities based on operational noncontextual ontic hidden variables. The nine equivalence classes of measurement and sources previously considered are once more investigated and their measurement and preparation noncontextuality consequences are derived (Eq (57) in~\cite{Krishna2017}). Once the constraints are obtained, noncontextual correlation polytopes, that is a characterisation of all the possible correlations in the noncontextual ontic model, are derived. From it, the facet inequalities that stem from this polytope are obtained. They are of the form (Eq (84) from~\cite{Krishna2017}):
\begin{equation}
    \sum_{i,j=1}^3 \alpha_{ij} \omega_{ij} \leq \beta
\end{equation}
with $\alpha_{ij}$ and $\beta$ integers, $\omega_{ij}$ being the operational correlations, regrouping the source and measurement assignments, that form the noncontextual polytope. By applying the noise map on the correlations, with $\omega{ij} = r^2$, a Peres-Mermin square robust inequality can be obtained~\cite{Krishna2017}:
\begin{equation}
    9 r^2 \leq 5
\end{equation}
which demonstrates contextuality provided $r \ge \frac{\sqrt{5}}{\sqrt{9}}$.

\subsubsection{Relation to the MKC controversery}
\label{sub:MKCOP}

In~\cite{Hermens2014}, an analysis is made of the relation between the MKC models and the noncontextual model operationally defined by Spekkens. The main divergence comes from the fact that in the operational model, every convex combination of preparation, transformation or measurement procedure must be represented by a convex sum of the corresponding probability measures. This highlights a deeper difference; in operational theory, two physical objects that can \textit{no longer} be distinguished (because some information have been lost for instance) are equivalent. Hermens argues that this metaphysical principle is too costly; in any case, the discussion helps to clarify the notion of contextuality underlied by Spekkens. 

\subsection{The contextual fraction and its limitations}
\label{sub:ContextualFraction}
We now turn to the formalism developed in section~\ref{Abramsky} of~\cite{Abramsky2017}. Consider an empirical model $e$. Assume it can be written according to the following convex decomposition
\begin{equation}
    e = \lambda e^{\text{NC}} + (1 - \lambda) e'
    \label{eq:contextualfraction}
\end{equation}
where $e^{\text{NC}}$ is a non-contextual model and $e'$ is another empirical model. Contextuality is quantified by the maximum possible value of $\lambda$ in such a decomposition. It is called the \textit{non-contextual fraction} of $e$. It is noted \textbf{NCF}($e$), and the \textit{contextual fraction} is \textbf{CF}($e$) = $1$ - \textbf{NCF}($e$). It was presented in~\cite{Mansfield2019}.

\par This decomposition has several advantages~\cite{Mansfield2019}:
\begin{itemize}
    \item It can be calculated using linear programming.
    \item It quantifies quantum-over-classical advantages in specific informatic tasks.
    \item It is a monotone with respect to the free operations of resource theories for contextuality.
\end{itemize}

It is an interesting alternative or maybe a complement to the hierarchy of contextuality, moving from a discrete to a continuous approach. However, this quantification does not take into account the no-signalling events. They are straightforwardly excluded by the compatibility principle in the ABS framework already exposed in Def.~\ref{def:compatibility}:
\begin{equation}
    \forall C, C' \ \in \mathcal{M}, \ e_ {C|C \cap C'} = e_ {C'|C \cap C'}.
\end{equation}
If we use the CbD terminology, the contextual fraction is limited to consistently connected systems~\cite{Kujala2019, Dzhafarov2020cbdABS}. In this sense, the sheaf-category framework is not yet able to treat the noisy experimental \textit{scenarii} that led to the considered Spekkens loophole. Possible insights to extend the model could be found in~\cite{Barbosa2014}. We leave a presentation of the PM square in this theory as a future work, once this problem has been treated.

\par In order to circumvent this limitation, we now turn to the CbD approach that is directly meant to take into account these signalling mistakes and their effects on contextuality. 

\subsection{The Peres-Mermin inequality in CbD theory}
\label{sub:CbDPM}

 We now write the scenario of the Peres-Mermin square from section~\ref{sub:PM} in the CbD language. We derive, in this case, new proper bounds of contextuality, and apply it to experimental results given in the Supplementary Material of~\cite{kirchmair2009}.

We follow the same protocol than the one presented in~\cite{Dzhafarov2015ThreeTypes} for other systems. A Peres-Mermin square system consist of $6$ triples of $9$ binary random variables that we usually wrote:

\begin{align}
    \mathfrak{S} = \{(A_{11}, A_{12}, A_{13}), (A_{21}, A_{22}, A_{23}), (A_{31}, A_{32}, A_{33}), \nonumber \\ (A_{11}, A_{21}, A_{31}), (A_{12}, A_{22}, A_{32}), (A_{13}, A_{23}, A_{33})  \}
\end{align}
However, in the CbD approach, we postulate that the same observables taken in two different contexts are \textit{a priori} not related, they are \textit{stochastically unrelated}.
We re-index the items of the Peres-Mermin items to be in line with the notions of Cbd. To do so, we now use the subscripts of $A_{ij}$ just to characterise the 9 obervables of the PM square, and indicate with an upper script the context in which the random variable is being measured. The system is now written
\begin{align}
     \mathfrak{S} = \{(A_{11}^1, A_{12}^1, A_{13}^1), (A_{21}^2, A_{22}^2, A_{23}^2), (A_{31}^3, A_{32}^3, A_{33}^3), \nonumber \\ (A_{11}^4, A_{21}^4, A_{31}^4), (A_{12}^5, A_{22}^5, A_{32}^5), (A_{13}^6, A_{23}^6, A_{33}^6)  \}
\end{align}
The connections are:
\begin{align}
    \mathfrak{C} &= \{ (A_{11}^1, A_{11}^4), (A_{12}^1, A_{12}^5), (A_{13}^1, A_{13}^6), (A_{21}^2, A_{21}^4), \nonumber \\
    &(A_{22}^2, A_{22}^5), (A_{23}^2, A_{23}^6), (A_{31}^3, A_{31}^4), (A_{32}^3, A_{32}^5), (A_{33}^3, A_{33}^6)  \}
\end{align}
In the ideal Peres-Mermin system, only the triples for which the last outcome is the product of the two previous outcomes can exist. Hence, $(-1, -1 -1)$, $(+1, -1, -1)$, $(-1, +1, -1)$, $(-1, -1, +1)$ and $(+1, +1, +1)$ are acceptable triples, but not the four other combinations. However, in the CbD analysis, we allow that
\begin{align}
    &\text{Pr}[A_{11}^1 = +1, A_{12}^1 = -1, A_{13}^1 = -1] \\
    & \text{Pr}[A_{11}^1 = +1, A_{12}^1 = -1, A_{13}^1 = -1] \\
    &\text{Pr}[A_{11}^1 = +1, A_{12}^1 = -1, A_{13}^1 = -1] \\
    & \text{Pr}[A_{11}^1 = +1, A_{12}^1 = -1, A_{13}^1 = -1] 
\end{align}
for instance. These are due to general noise, signalling or others, and they do appear in experimental results. We consider couplings $S$ for our connections, denoted
\begin{equation}
 S = \{  \{ A_{ij}^k*, A_{ij}^l*\} : \{i,j\} \in \{1;2;3\} , \{k,l\} \in \{1;2;3;4;5;6\}, k \neq l\}
\end{equation}
We can now evaluate the degree of contextuality of our system $(\mathfrak{S},\mathfrak{C})$ with the formula:
\begin{equation}
   \text{CNTX}(\mathfrak{S},\mathfrak{C}) = \Delta_{\text{min}}(\mathfrak{S},\mathfrak{C}) - \Delta_0(\mathfrak{C})
\end{equation}
where
\begin{equation}
    \Delta_0(\mathfrak{C}) = \frac{1}{2} \sum_{i,j} |\moy{A_{ij}^k*}-\moy{A_{ij}^l*}|
\end{equation}
and
\begin{equation}
    \Delta_{\text{min}}(\mathfrak{S},\mathfrak{C}) = \min_{S \text{ for } \mathfrak{C}} \sum_{i,j} \text{Pr} [A_{ij}^k* \neq A_{ij}^l*].
\end{equation}
We compute, according to the method established in~\cite{Dzhafaro2014} and~\cite{Dzhafarov2015}, that:
\begin{align}
     \Delta_0(\mathfrak{C}) = \frac{1}{2} &( 0.033+0.031+0.03+0.004 \nonumber \\
      &+0.086+0.02+0.022+0.093+0.013 ) \nonumber \\
      &= 0.166
\end{align}
since
\begin{align}
    &\moy{A_{11}^1} = 0.031 \\
    &\moy{A_{11}^4} = 0.064 \\
    &\moy{A_{12}^1} = 0.0 \\
    &\moy{A_{12}^5} = -0.031 \\
    &\moy{A_{13}^1} = -0.92 \\
    &\moy{A_{13}^6} = -0.95 \\
    &\moy{A_{21}^2} = -0.013 \\
    &\moy{A_{21}^4} = -0.009 \\
    &\moy{A_{22}^2} = 0.017 \\
    &\moy{A_{22}^5} = -0.069 \\
    &\moy{A_{23}^2} = -0.938 \\
    &\moy{A_{23}^6} = -0.918 \\
    &\moy{A_{31}^3} = -0.027 \\
    &\moy{A_{31}^4} = -0.005 \\
    &\moy{A_{32}^3} = -0.007 \\
    &\moy{A_{32}^5} = -0.1 \\
    &\moy{A_{33}^3} = -0.88 \\
    &\moy{A_{33}^6} = -0.893 
\end{align}

where we have taken the correspondence:
\begin{align}
    &A_{11}^1 \rightarrow \sigma_z^{(1)} \\
    &A_{11}^4 \rightarrow \sigma_z^{(1)}   \\
    &A_{12}^1 \rightarrow \sigma_z^{(2)} \\
    &A_{12}^5 \rightarrow \sigma_z^{(2)}   \\
    &A_{13}^1 \rightarrow \sigma_z^{(1)} \otimes \sigma_z^{(2)} \\
    &A_{13}^5 \rightarrow \sigma_z^{(1)} \otimes \sigma_z^{(2)}  \\
    &A_{21} \rightarrow \sigma_x^{(2)} \\
    &A_{22} \rightarrow \sigma_x^{(1)} \\
    &A_{23} \rightarrow \sigma_x^{(1)} \otimes \sigma_x^{(2)} \\
    &A_{31} \rightarrow \sigma_z^{(1)} \otimes \sigma_x^{(2)} \\
    &A_{32} \rightarrow \sigma_x^{(1)} \otimes \sigma_z^{(2)}
\end{align}
and the context given by the exponant corresponds to the column. Note that there is a mistake in the tabular of the Supplementary information, in the third line of the second column it must be $\sigma_x^{(1)} \otimes \sigma_z^{(2)}$ instead of  $\sigma_x^{(1)} \otimes \sigma_x^{(2)}$.

Now, 
\begin{align}
    &a_1 = \moy{A_{11}^1 A_{12}^1 A_{13}^1} = 0.924 \\
    &a_2 = \moy{A_{21}^2 A_{22}^2, A_{23}^2} = 0.931\\
    &a_3 = \moy{A_{31}^3 A_{32}^3, A_{33}^3} = 0.900 \\ 
    &a_4 = \moy{A_{11}^4 A_{21}^4 A_{31}^4} = 0.900\\
    &a_5 = \moy{A_{12}^5 A_{22}^5 A_{32}^5} = 0.895 \\
    &a_6 = \moy{A_{13}^6 A_{23}^6 A_{33}^6} = -0.913
\end{align}
so, with the notations and results from~\cite{Dzhafarov2015ThreeTypes}, that is:
\begin{equation}
    s_{\text{odd}}(a_1,\cdots,a_n) = \max_{\text{odd number of -'s}}\sum_{i=1}^n (\pm a_i) 
\end{equation}
and
\begin{equation}
     \Delta_{\text{min}}(\mathfrak{S},\mathfrak{C}) =\frac{1}{2} \max(2 \Delta_0(\mathfrak{C})  , s_{\text{odd}}(a_1,\cdots,a_6)-4)
\end{equation}
we have
\begin{equation}
    s_{\text{odd}}(0.924,0.931,0.900,0.900,0.895,-0.913) = 5.463
\end{equation}
\begin{equation}
     \Delta_{\text{min}}(\mathfrak{S},\mathfrak{C}) = \frac{1}{2} \max \left(0.332,5.463-4 \right) = 0.732
\end{equation}
And since:
\begin{equation}
    \Delta_{\text{min}}(\mathfrak{S},\mathfrak{C}) = 0.732 \geq 0.166 = \Delta_0(\mathfrak{C})
\end{equation}
we have proven that the experiment of~\cite{kirchmair2009} has indeed witnessed contextuality, even after having taken into account the signalling mistakes.
We can calculate the degree of contextuality in the sense of~\cite{Dzhafarov2015}:
\begin{equation}
    \text{CNTX}(\mathfrak{S},\mathfrak{C}) = 0.566
\end{equation}
Hence, the CbD approach enabled us to present a quantification of the Peres-Mermin contextuality inequality, able to cope with the numerous error terms of an experimental realisation, and notably with the signalling errors. Let us note that this technique is not able to differentiate the different sources of noise. We applied this model to experimental data and confirmed the presence of contextuality in this case, with this new approach. 

\subsection{Conclusion on the derivation of new bounds}
\label{sub:CclNB}

With the four previously considered models, we can draw the following conclusion. The MKC loophole is closed in CSW but without experimental certification, closed in operational theory at the cost of a (very affordable for us) metaphysical price, and it is unknown for the sheaf category. The Outcome Determinism issue is in the same way tackled in the CSW framework, and may be closed by relaxing Eq.~\eqref{eq:Sheafcompatibility}. In the CbD framework, outcome undeterminism is immediately assumed. We consider that the MKC loophole is also closed since CbD builds contextuality directly from experimental data, without referring to the quantum world. 

\vspace{0.5cm}

\par We now turn to the second part of our program, related to the recovery of empirical data within a quantum theory model which takes noise into account. To do so, we will consider how realistic measurements can be modelised with the help of the formalism of POVMs and how quantum noise channels can model the evolution of states.


\section{A quantum theoretical model that takes noise into account} 
\label{NoiseforPM}

\par In the experiments we consider, there exist many sources of noise~\cite{kirchmair2009, CabelloConv}, that can be due to detection apparatus imperfections, misalignments, to the limited number of data we can obtain, among others. To describe realistic experiments, we need to abandon the ideal case of the Peres-Mermin square in the quantum theory approach. The presence of signalling had force us to derive new bounds, in the previous part, but it an also, along with compatibility issues, lead to an impossibility to witness a violation. We therefore treat, in the quantum framework, the amount of noise that is acceptable. 

\par We first show that for some class of sufficiently noisy POVMs, the relation of compatibility between the observables of the Peres-Mermin square is preserved in subsection~\ref{subsec:MotherPOVM}. However, it is likely that the cost of this amount of noise will be to prevent us from witnessing contextuality. Since we would like to model realistic experiments, we try to take into account the incompatibility of measurements in the laboratory. To do so, we give a brief presentation of the usual quantum channels that model noise in experiments in subsection~\ref{sub:noisychannels}, and then show their application on the Peres-Mermin inequality in subsection~\ref{sub:Implementationl}, relying on Szangolies'work~\cite{Szangolies2015}. We conclude in subsection~\ref{sub:Adhoccorrections} by reviewing the different attempts that have been made to derive new noncontextual bounds that are robust to noise, by the addition of \textit{ad-hoc} error terms, and expose their limits.

\subsection{A simple noise model with compatible observables}
\label{subsec:MotherPOVM}

We replace the projective measurements represented by Pauli operators on the quantum version of the Peres-Mermin square by POVMs. The notion of compatibility between observables is no longer reducible to the commutation of observables. POVMs are said to be jointly measurable when they are obtained by marginalisation from a mother POVM. The degree of compatibility of noisy observables can be quantified, according to the noise models considered. Different ways to proceed have been exposed and compared, notably in ~\cite{Cavalcanti2016} and a global and unifying approach has recently been given in~\cite{Designolle2019}.

\par We use the results of~\cite{Rao2020}. Recalling that the Pauli operators, which are optimal measurements since their eigenbases are mutually unbiased~\cite{Rao2020,Kurzynski2010}, are based on the spectral decomposition:
\begin{equation}
    \sigma_i = \sum_{x_i \in \{-1;1\}} x_i P(x_i) \ \forall i \in \{1,2,3\}
    \label{eq:SDPauli}
\end{equation}
where $x_i$ are the eigenvalues of $\sigma_i$ and $P(x_i)$ the projectors corresponding to the associated eigenstate.  Consequently,
\begin{equation}
    P(x_i) = \frac{1}{2} (\mathds{1} + x_i \sigma_i)
    \label{eq:ProjPauli}
\end{equation}
The probabilities associated with each projectors are 
\begin{equation}
    p(x_i) = \Tr\left( \rho(\theta) P(x_i) \right) = \frac{1}{2} (1 + x_i \theta_i)
    \label{eq:ProbProjPauli}
\end{equation}
with the requirement $-1 \leq \theta_i \leq 1$. Now, we turn them into unsharp projectors by introducing a noise parameter $\eta \in [0,1]$, such that
\begin{equation}
    E(x_i) = \frac{1}{2} (\mathds{1} + \eta x_i \sigma_i)
    \label{eq:ProjPauliNoise}
\end{equation}

It  was  shown  in  \cite{Busch1986}  that  the  operators $E:=\{E(x_1),E(x_2),E(x_3)\}$ are jointly measurable with global POVM $G=\{G(x_1,x_2,x_3)\}$
\begin{equation}
    G(x_1,x_2,x_3) = \frac{1}{8} \left( \mathds{1} + \eta \left( x_1 \sigma_1 + x_2 \sigma_2 + x_3 \sigma_3 \right) \right)
    \label{eq:POVMG}
\end{equation} when $\eta \in [0, 1/\sqrt{3}]$ only. 
\par Hence, this gives us a sufficient condition to establish a new Peres-Mermin inequality with jointly measurable unsharp projectors. This is not a necessary condition because we have imposed that observables on each side of the tensor product are compatible, which is for instance not the case in the original version. Note however that any kind of noise added in the original version breaks the commutation relation. 
\par The noise model proposed above covers interesting experimental situations. A generalisation could be however envisaged, in which all observables would take the following form:
\begin{equation}
    \forall i \in \{1,2,3\}, \tilde{\sigma_i} = \sigma_i + \eta \mathds{1} + \eta_j \sigma_j + \eta_k \sigma_k
    \label{eq:GenNoiseModelPauli}
\end{equation}
where $\{j,k\} \neq i$ with noise parameters $\eta, \eta_j, \eta_k$ to be determined, in order for the observables to be jointly measurable. However, since we want to modelise cases in which noise is not as strong as it is required here to obtain jointly measurable observables, we will consider more general modelisations. We note that these ideas have been already developed and that noncontextual inequalities have been obtained through them~\cite{Liang2011} in the operational framework (see notably part VII). They have even been implemented in experiments~\cite{Zhan2017}.

\subsection{Presentation of general noisy channels}
\label{sub:noisychannels}

 There exist several noise models whose purpose are to take into account general type of errors in physical setups. Each one can be expressed through a transformation, characterised by Kraus Operators. We shortly describe them according to~\cite{PreskyllChap}, and then represent their effects on the Peres-Mermin inequality on a graph.

\subsubsection{Depolarising noise}
A depolarising channel corresponds to a physical process for which the system remains intact with probability $p$ and an error occurs with probability $1-p$. The error, in the case of a qubit, can be a bit flip, a phase flip or both, with equal probability. Hence, the Kraus operators that modelise this process are given by 
\begin{align*}
    E_0 = \sqrt{1 - \frac{3p}{4}} \mathds{1} \nonumber \\
    E_1 = \sqrt{\frac{p}{4}} \sigma_x \nonumber \\
    E_2 = \sqrt{\frac{p}{4}} \sigma_y \nonumber \\
    E_3 = \sqrt{\frac{p}{4}} \sigma_z \nonumber \\
    \label{eq:DepNoiseKraus}
\end{align*}
The transformation can be written, accordingly, as
\begin{equation}
    \mathcal{E} (\rho) = \frac{p}{4} \left( \sigma_x \rho \sigma_x + \sigma_y \rho \sigma_y + \sigma_z \rho \sigma_z \right)  + \left( 1 - \frac{3 p}{4}\right) \rho,
    \label{eq:DepNoise}
\end{equation} or in the other form
\begin{equation}
    \mathcal{E} (\rho) = p \frac{\mathds{1}}{Tr(\mathds{1})} + (1 - p) \rho,
    \label{eq:DepNoise2}
\end{equation}
which expresses directly that this transformation results in combining the initial state with the completely mixed state. We can infer from this general model of noise a bit flip or a phase flip model, which unbalances the probability of each error terms, according to our purpose.

\subsubsection{Bit flipping}
This noise model describes to probability for a state to be flipped into an orthogonal state. Its Kraus operators are thus:
\begin{align*}
    E_0 = \sqrt{1 - p} \mathds{1} \nonumber \\
    E_1 = \sqrt{p} \sigma_x \nonumber 
    \label{eq:BitFlipNoiseKraus}
\end{align*}
and the transformation can be written:
\begin{equation}
     \mathcal{E} (\rho) =P \left( \sigma_x \rho \sigma_x \right)  + \left( 1 - p \right) \rho
    \label{eq:BitFlipnoise}
\end{equation}

\subsubsection{Phase Damping}
This noise model describes a process of decoherence, the loss of phase information. Its Kraus operators are thus:
\begin{align*}
    E_0 = \sqrt{1 - \frac{p}{2}} \mathds{1} \nonumber \\
    E_1 = \sqrt{\frac{p}{2}} \sigma_z \nonumber 
    \label{eq:PhasDampNoiseKraus}
\end{align*}
and the transformation can be written:
\begin{equation}
     \mathcal{E} (\rho) =\pmat{\rho_{00}}{\sqrt{1-p}\rho_{01}}{\sqrt{1-p}\rho_{10}}{\rho_{11}}
    \label{eq:PhasDampnoise}
\end{equation}
where we see that the off-diagonal terms are the only ones to decohere.

\subsubsection{Amplitude damping noise}
The amplitude damping channel modelises the  decay of an excited state of a two-level atom, due to the spontaneous emission of a photon, with a certain probability. Its Kraus operators are:
\begin{align*}
    E_0 = \pmat{1}{0}{0}{\sqrt{1-p}} \nonumber \\
    E_1 = \pmat{0}{\sqrt{p}}{0}{0} \nonumber \\
    \label{eq:AmpNoiseKraus}
\end{align*}
and the transformation is 
\begin{equation}
    \mathcal{E} (\rho) = \pmat{\rho_{00}+p\rho_{11}}{\sqrt{1-p}\rho_{01}}{\sqrt{1-p}\rho_{10}}{(1-p)\rho_{11}}
    \label{eq:AmpNoise}
\end{equation}

\subsection{Implementation of noisy channels on the Peres-Mermin Square experiment}
\label{sub:Implementationl}

\subsubsection{Applying noise on sequential measurements}
\label{subsub:noiseseq}
Although the expectation value of an observable $A$ measured on a state $\rho$ under the effect of noise modelised by a channel $\mathcal{E}$ is $\moy{A} = \Tr \left( A \mathcal{E} (\rho)\right)$, the expression for a sequence of incompatible measurements is a little bit more complex. We derive it according to~\cite{Guhne2014} and~\cite{Szangolies2015}. In our case, the observables can yield two outcomes: $\pm 1$. If we record the outcome $-1$ on the first measurement, the state becomes:
\begin{equation}
    \rho_{A^-} = \frac{\Pi_A^{-} \rho \Pi_A^{-}}{\Tr \left( \Pi_A^- \rho \right)}
    \label{eq:MeasurementProcess}
\end{equation}
where $\Pi_A^-$ is the projector on the eigenspace of $A$ associated with the eigenvalue $-1$, according to Lüders's rule.
The probability of measuring $1$ for the second observable $B$ is then 
\begin{equation}
    \text{Pr}(B=1|A =-1) = \Tr \left( \Pi_B^+ \frac{\Pi_A^- \rho \Pi_A^-}{\Tr(\Pi_A^- \rho)} \right)
    \label{eq:ProbaAB}.
\end{equation}
The expectation value of the succession of observables A and B is 
\begin{align}
    \moy{AB} &= \text{Pr}(B=1|A=1)-\text{Pr}(B=-1|A=1) \nonumber \\
    &- \text{Pr}(B=-1|A=1)+\text{Pr}(B=-1|A=-1).
    \label{eq:MeanAB}
\end{align}
After some calculations and simplifications, we can obtain the expression:
\begin{align}
    \moy{ABC} &= \Tr \left( C \mathcal{E} \{ \Pi_B^+  \mathcal{E} ( \Pi_A^+ \rho \Pi_A^+ - \Pi_A^- \rho \Pi_A^-) \Pi_B^+  \right.\nonumber \\
    &- \left. \Pi_B^-  \mathcal{E} ( \Pi_A^+ \rho \Pi_A^+ - \Pi_A^- \rho \Pi_A^-) \Pi_B^- \} \right)
    \label{eq:MeanABC}
\end{align}
for a sequential noisy measurement of the observables $ABC$, where the noise model is given by $\mathcal{E}$. For some thoughts on the degree of generality of the quantum noise models, see Appendix.~\ref{sub:dualitystatemeasurement}

\subsubsection{Effect of noise models on the Peres-Mermin inequality}

In agreement with~\cite{Szangolies2015}, the different noise models applied on the Peres-Mermin square lead to the following expressions:
\begin{align}
    &\moy{\chi_{PM}}^{DN} = 6 (p-1)^2 \\
    &\moy{\chi_{PM}}^{BF} = 6 - 28p + 56p^2 - 48p^3 + 16p^4 \\
    &\moy{\chi_{PM}}^{AD} = (1-p)(2+4\sqrt{1-p}-(4+3\sqrt{1-p})p \\ 
    &                         +6p^2-2p^3) \nonumber
\end{align}
\begin{align}
\moy{\chi_{PM}}^{PD} &= \frac{1}{2}\left( 4+8\sqrt{1 - p}-4(2 + 3\sqrt{1-p})p \right. \\
                     &+ (15 + 16\sqrt{1-p})p^2 - (21 + 16\sqrt{1-p})p^3 \nonumber \\
                     &+ 2(11 + 6\sqrt{1-p})p^4 -(17 + 6\sqrt{1-p})p^5 \nonumber \\
                     &+ \left. 2(5 + \sqrt{1-p})p^6 - 4p^7 + p^8 \right) \nonumber
\end{align}

\begin{figure}[h]
    \centering
    \includegraphics[width=1.\linewidth]{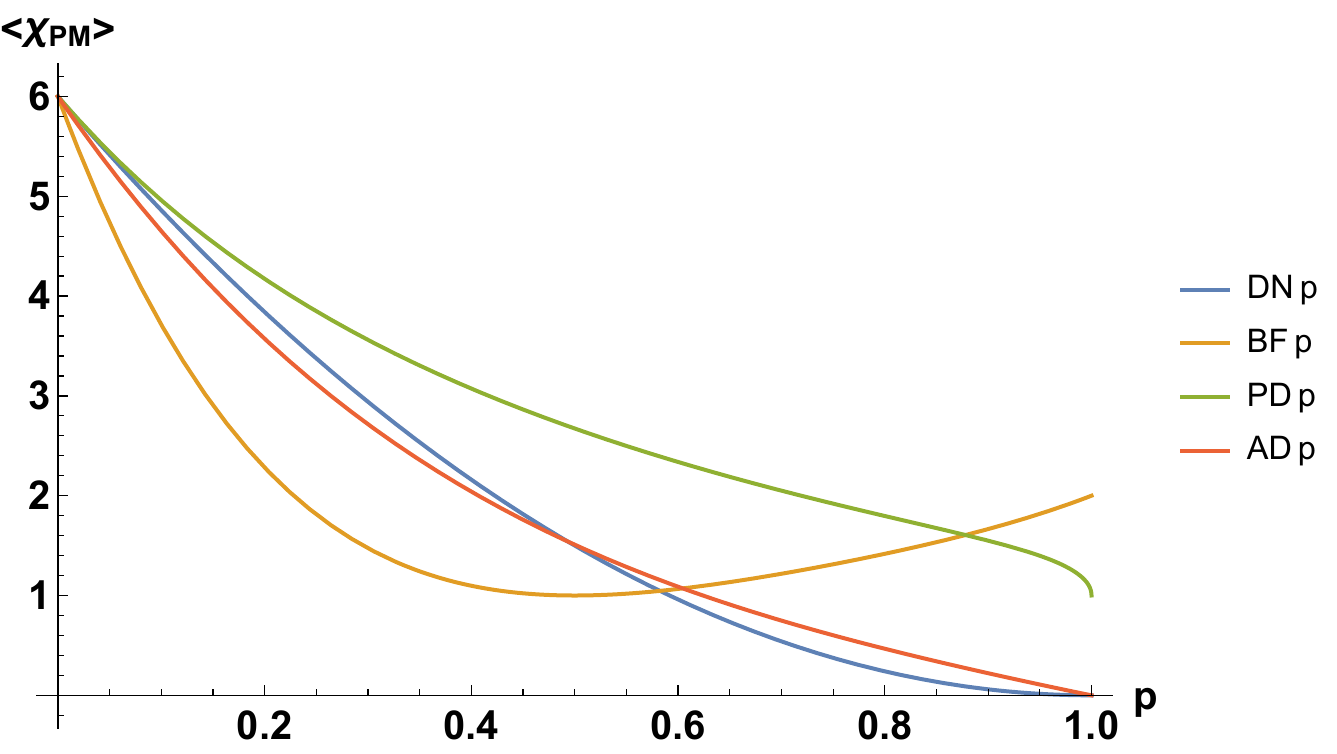}
    \caption{Evolution of $\moy{\chi_{PM}}$ for different models of noise. Depolarisation Noise in blue, Bit Flipping in orange, Phase Damping in green, Amplitude Damping in red. In agreement with Figure 3.16 from~\cite{Szangolies2015}.}
    \label{fig:NoisePM}
\end{figure}

They are represented on Fig.~\ref{fig:NoisePM}. Let us mention once more that this graph is not suited to prove contextuality, but merely to show that the experimental results are not incompatible with the quantum theory. The main lesson we can take for this graph is that all sources of noise have an important impact on the final values we get, Bit Flipping being nearly twice as destructive as the three other models for small values of noise. This is understandable in the sense that any sign flip reverses the sign of one terms of the inequality. For instance, it turns a maximally contextual result ($\moy{S} = 6$) in a noncontextual result ($\moy{S} = 4$) in the ideal case. 

\subsection{\textit{Ad-hoc} corrections for experiments}
\label{sub:Adhoccorrections}

These results yield a possible explanation to the experimental results we get, consistent with the quantum theory. However, on the other side of the inequality, the classical bound must also take into account the error terms. We review how noise, notably compatibility issues, have been addressed, and conclude on the necessity to use the CbD protocol. Hence, we disconnect the certification of the presence of contextuality in an experiment with the predictions of results due to quantum theory. 

\subsubsection{Error terms corrections due to non-compatibility}
\label{subsub:Errorterms}

Several works have tried to directly take into account the possible experimental mistakes. In~\cite{Guhne2014}, the probability that the first measurement flips the sign of the second is expressed with the terms $p^{\text{flipp}}[AB]$, according to:
\begin{align}
    p^{\text{flipp}}[AB] &= p[(B_1^+|B_1) \ \text{and} \ ((B_2^-|A_1 B_2)] \\
    &+ p[(B_1^-|B_1)\ \text{and} \ ((B_2^+|A_1 B_2)]. \nonumber
\end{align}
This example was extended to the case of three measurements in a row, and the KS-noncontextual inequalities were derived again to take into account all these additional terms.

\par However, these probabilities are not experimentally accessible since $B$ can not be simultaneously measured in the two different places in a same sequence. They were then upper-bounded and turned them into three kinds of new inequalities with experimentally accessible terms~\cite{Guhne2014, Guhne2010}. The first kind used a principle of \textit{cumulative noise}, assuming that the more measurements were made the more disturbance would occur to the system. The second kind corrects the inequalities according to an estimation of the incompatibility of observables. It presents tests of compatibility for observables, and study the ratio of experimental events that pass these tests. The third kind of correction is an estimation and a quantification on the compatibility of observables that is in its form rather similar to the one of~\cite{Winter2014} for contextuality.

\par Yet, it was shown in~\cite{Szangolies2013} that a hidden-variable model, for which the variables evolve at each measurement but independently from any context, could maximally violate the first kind of these new inequalities. Szangolies then suggested to use Leggett and Garg ideas and to reorder the sequences of observables used in the noncontextual inequalities. Each observable should be measured at the same place in different contexts. The Peres-Mermin inequality of inequation~\eqref{eq:PM} thus becomes:
\begin{align}
\moy{S} &= \moy{A_{11}A_{12}A_{13}} + \moy{A_{23}A_{21}A_{22}} +  \moy{A_{32}A_{33}A_{31}} \nonumber \\  &+\moy{A_{11}A_{21}A_{31}} +  \moy{A_{32}A_{12}A_{22}} - \moy{A_{23}A_{33}A_{13}}.
    \label{eq:PMordered}
\end{align}
It was proven in~\cite{Szangolies2013} that this reordering, while not closing the compatibility loophole, significantly improves the inequality by ruling out large class of hidden-variable models that would otherwise violate the inequality. 

\subsubsection{Limits of error correction terms and necessity to adopt a CbD approach}
\label{subsub:BeyondSeq}
Although Cabello \textit{et al.} argued in~\cite{Cabello2016} that the contextuality assumptions needed for a sequential measurement sequence are equally well-motivated than for a joint measurement setting, and that sequential settings are actually simpler to realise experimentally, they still suffer from two main flaws. The first one is that typical projective measurements on systems are demolition measurements that either prevent the Lüders's rule to apply or simply absorb the system and prevent other measurements to be made~\cite{Cabello2016}. The second one is that sequential measurements always yield deviations due to signalling terms. This is why corrections terms, such as those from subsection~\ref{sub:Winter}, or a complete change of point of view, as the CbD approach described in section~\ref{Cbd}, was needed. 

\subsubsection{CSW is fully compatible with CbD}
\label{subsub:CSWWinter}

Cabello showed that the CSW graph theory we presented in section~\ref{CSW} could be used to transform any kind of noncontextuality inequalities into new ones, involving only two measurements, and for which the work Winter from~\cite{Winter2014} and the CbD approach~\cite{Kujala2015} is well-defined. Indeed, it was shown that any noncontextuality inequalities can be represented by a graph. Now, given a graph $G$ with vertex set $V(G)$ and edge set $E(G)$, it is proven that 
\begin{equation}
    S = \sum_{i \in V(G)} P(1|i) - \sum_{(i,j) \in E(G)} P(1,1|i,j)
\end{equation}
is still tightly bounded by the independence number $\alpha(G)$ for noncontextual hidden-variable models and by the Lov\'asz number $\mathcal{V}(G)$ for quantum theory, where $P(1|i)$ is the probability of obtaining $1$ when observable $i$ is measured, and $P(1,1|i,j)$ is the joint probability of obtaining result $1$ for $i$ and result $1$ for $j$.

\section{Conclusion on the meaning of contextuality experiments}
\label{FinalProtocol}
We sum-up the discussion of this part. We present a protocol that takes into account the discussions of the last parts in~\ref{sub:Protocol}, conclude on the relevance and scope of the different approaches in section~\ref{sub:Ccl} and open our work on a quantum foundation perspectives in~\ref{sub:qf}.

\subsection{Setting a protocol to derive experimentally robust inequalities}
\label{sub:Protocol}
An experimental protocol of KS-contextuality could take the following form
\begin{enumerate}
    \item Choose a theoretical inequality~\cite{CSW2014,Acin2015}
    \item Reorder the term lower the compatibility loopholes~\cite{Szangolies2013}
    \item Analyse the robustness to noise of the inequality
    \item Transform into a two-point correlation inequality
    \item Use CbD 2.0 to estimate the amount of contextuality
\end{enumerate}

\subsection{What we can learn from our approach to KS-contextuality experiments}
\label{sub:Ccl}

We analysed in this part the difficulties that arise when one wishes to adapt the non-contextuality inequalities derived from KS-contextuality into experimentally robust ones. We tried to show how the criticism of MKC and Spekkens (section~\ref{Analysis}) can be answered, inside the quantum theory (\ref{sub:Winter}) and in a more global scope (\ref{sub:CbDPM}) to derive robust bounds. We then presented protocols to retrieve the empirical results within quantum mechanics (section~\ref{NoiseforPM}). The linking of these approaches yields
\begin{enumerate}
    \item a protocol to establish non-contextual inequalities that is directly applicable to imperfect sets of measurements - a violation of which shows that “Nature is contextual”,
    \item a fully compatible QM model that reproduces our correlation, and thus shows that “Quantum theory has not been proven to be false with these experiments".
\end{enumerate}

\par It is widely believed that the first point is of larger scope, \cite{Spekkens2005, CabelloConv, Dzhafarov2015, Dzhafarov2016}, and we absolutely agree with it. However, it seemed to us that the second might be of interest as well. First, quantum physics is, to the best of our knowledge, the first natural science theory that needs to present a contextual description. Second, it would have been rather undermining for the scope of the theory that it could not be extended in this case to experimental situations. Third, we believe that inside QM it can be interesting to disconnect and quantify, in statistical outcomes, how the quantifiable resource of contextuality is affected by incompatibility and signalling.

\subsection{Can contextuality be a theory-independent concept?}
\label{sub:philocontest}

In~\cite{Hermens2014}, Hermens explicitly states the paradoxical aspect of contextuality: if we assume that an object is defined by a certain identity, or stability, then we can not say that this is the same object if it behaves differently (in difference contexts). In other words, if our probability distribution is the same, there is no contextuality, and if it is a different one, then it is a different object. It is our opinion that CbD answers adequately this criticism, by assuming another theoretical layer above the directly observed probability outcomes. Namely, it purposefully choses to study as relevant scientific objects mathematical connections, although they are not directly derivable in the QM theory, or even directly observed. It differs from the operational view in which there is a strict equivalence form which they stem. In~\cite{Szabo2020a}, it is argued that the requirement for ontic models to be defined according to a strict equivalence between measurements and self-adjoint operatators (which is the case in all operational frameworks) comes more from the QM representation than from an independent an intuitive concept of contextuality. Indeed, the author states a concept of contextuality where (i) each operator is uniquely associated with a measurement, and where (ii) commuting operators represent simultaneous measurements. It is shown that (i) is logically different from the definition of contextuality used in~\cite{Spekkens2005}, and is reminded that commutativity has no reason to imply simultaneous measurability in the quantum framework. This clarification is supported by an analysis of the Peres-Mermin square scenario, where three hidden variable models reproduce QM results if they violate (i), (ii), or both conditions~\cite{Szabo2020b}.

\par But then, what meaning can we give to contextuality? If we want to make of it a metaphysical principle, it should be defined independently of the theories on which it is applied. For Hermens, this is not really possible, neither from the operational theory (as we saw in~\ref{sub:MKCOP}), nor from the initial formulation def~\ref{def:NCHermens}, too rooted in the quantum framework~\cite{Hermens2014}.

\par In~\cite{Deronde2016}, a distinction is made clear between two different notions of contextuality. One epistemic definition, based on measurement outcomes and statistics, and one ontic definition based on the formalism of projection operators, which is the KS theorem. According to~\cite{Deronde2016}, one should not try to give an epistemic reading to the KS theorem (in terms of outcomes of observables, because they change according to the context since they can not be measured simultaneously),  because of the destructive nature of measurements in QM and to the no-cloning theorem. In that sense, proving experimentally the KS theorem does not even make sense, since the KS theorem is here to explain the structure of QM. This is why the noncontextual inequalities are fundamentally different. In the wake of the quantification of contextuality that is nowadays underway, it seems that they can not only serve to rule out other (meta)physical theories, but that, by highlighting the private area of quantum mechanics, they provide a material resource theory~\cite{Amaral2019}.

\subsection{Towards a Quantum Foundations perspective}
\label{sub:qf}
If we consider the four structural theories considered, one is directly rooted in the framework of quantum mechanics, the graph theoretic approach of CSW, two are defined operationally, Spekkens operational framework and CbD, and the sheaf structure is defined on a mathematical meta-level (with regard to the quantum theory). However, the quantum theory is contextual, as proven by the KS-theorem. Besides, operational framework, sheaf category and CbD are able to confront noncontextual hidden variable models, but at the price of being defined outside the quantum framework. 
\par In this sense, we can say that, at the level of natural phenomena for which the quantum theory is efficient, nature has been shown to be contextual (with all the limitations to this formulation seen in~\ref{sub:MKC} and~\ref{sub:philocontest}). Besides, quantum theory has not been proved to be wrong yet. Hence, a  possibly interesting work would be to compare the most general prevision theory, based on a probabilistic framework \textit{and} contextual, to quantum mechanics. Works in that direction can be found in~\cite{Svozil2020, Bitbol2014, deRonde2015}. We acknowledge that the reconstruction of quantum theory from a minimal set of physics principle has already been much-studied field, as evidenced by the general perspectives derived in~\cite{Fuchs2009, Grinbaum2006, Wilce2017, Janotta2014,Pitowsky2006}, sometimes even from a contextuality point of view~\cite{Henson2015}. Other works have shown that no extension of quantum theory can yield more accurate prediction of outcomes, if it is build on a probability basis~\cite{Colbeck2011, Colbeck2016} (see also~\cite{Hermens2020} for a critical review).
\par It might be of interest to investigate the extension of the property of contextuality in microscopic nature. Indeed, the quantum to classical transition has been efficiently modelized by decoherence(~\cite{Zurek2006}), but some argued that it was unsatisfactory from an epistemological point of view~\cite{deRonde2016Decoherence} (that it was rather an ad-hoc mecanism than an explanation). This transition has also recently been studied from an information theory perspective~\cite{Czekaj2017}. If our last assumption (that quantum mechanics is the only general theory of prevision based on probabilities and contextuality) is correct, we could be able to witness this transition as a process through which the objects of physical theories cease to behave in a contextual fashion. But for which reason?

\clearpage

\section{General conclusion}
\label{UseofContextuality}

\subsection{Sum-up}

In this review, we showed how contextuality has risen from a discrete logical problem to a quantifiable quantum quantity. To do so, we first made a review of the major steps of contextuality, from its apparition through philosophical and logical problems to its adaptation into a mathematical formalism. We chose to present the major frameworks that have been developed to capture the notion of contextuality, and we tried to show the links between them. To finish, we studied the experimental tests that have been proposed, discussed the presence of loopholes and the relevance of the protocols aimed to close them. We argued that the sources of noise should be classified according to the modifications they imply on the proofs of contextuality. We emphasized that some experiments are tests against noncontextuality, while others estimate an amount of contextuality from quantum mechanics.

\par The major part of this work has been bibliographical. We believe that its value comes from the following points:
\begin{enumerate}
\item Far from being exhaustive, it is nevertheless an introduction to the wide scope field of contextuality.
\item It provides some links between the major theoretical works of contextuality.
\item It regroups and discusses the major controversies concerning the experimental certification of contextuality.
\item It provides a robust protocol completed with calculations for the specific case of the Peres-Mermin square experiment. 
\end{enumerate}

\subsection{Experimental realisations}
\label{sub:ExperimentalRealisations}
The different approaches we presented so far have been experimentally tested. Photons are the most popular information carrier for these correlation experiments. We list some of these realisations. The Spekkens approach has been implemented and tested by Mazurek \textit{et al.}~\cite{Mazurek2017, Mazurek2016} and Zhan \textit{et al.}~\cite{Zhan2017}. Cabello and Gühne have interpreted numerous experiments~\cite{Guhne2017conf} of KS-contextuality, from the inital Peres-Mermin square tested with sequential measurements~\cite{Cabello2008, DAmbrosio2012} to the one performed by Marques \textit{et al.} on KCBS scenario~\cite{Marques2014} which takes into account most of the criticism seen in this part, such as the modification of the upper bound of the hidden-variable model. A comparison between the capacity of single photons or coherent states to violate noncontextual inequality was performed in~\cite{Zhang2019}. For a more exhaustive review, the reader may refer to~\cite{Budroni2021}.

\par Let us give a word about the quest to obtain the best precision when measuring incompatible observables. We recall the three different messages of the Heisenberg Uncertainty Principle~\cite{Ringbauer2014}:
\begin{itemize}
    \item “A system can not be prepared such that a pair of non-commuting observables are arbitrarily well defined.
    \item Such a pair of observables cannot be jointly measured with arbitrary accuracy.
    \item Measuring one of these observables to a given accuracy disturbs the other accordingly."
\end{itemize}  
Generalisations of the Heisenberg Principle have been derived to quantify the precision with which two observables could be known, in the joint and measurement disturbed scenario~\cite{Branciard2013}. They take into account the tradeoff between accuracy and disturbance of observables~\cite{Branciard2014}. They were experimentally tested~\cite{Ringbauer2014, Kaneda2014}, in a quest to test the ultimate measurement uncertainty limits.

\subsection{Quantum computation advantage and a little bit more}
To conclude, contextuality has been identified as an advantage in Quantum Information, just like entanglement, squeezing or non locality, in the wake of the great Q.I surge, passing from “this property is weird" to “how can this property be useful?". Let us mention, without being exhaustive, that the theory of magical state distillation made of contextuality a resource for universal quantum computation \cite{Karanjai2018, Karanjai2017conf, Howard2014, Delfosse2015} and for measurement-based quantum computation~\cite{Raussendorf2013}, as well as quantum communication~\cite{Saha2019}. The Sheaf-theoretic aspect has been studied as a source of quantum advantage in~\cite{Mansfield2018ressource}. Contextuality has been used for cryptography \cite{Eckert1991, Bruckner2004, Colbeck2009, Barrett2005} and reduction of communication complexity \cite{Bruckner2004}. In fact, as we saw with the Spekkens approach and the CbD one, contextuality has been shown to go beyond the quantum theory. Several publications have shown the relations between the sheaf-theoretic approach of contextuality and phenomena such as relational database theory, robust constraint satisfaction, natural language semantics and logical paradoxes, a general theory of which, called \textit{contextual semantics}, is given in~\cite{Abramsky2019general}. Finally, the deeper and more general work in that direction is the formalisation of contextuality as a ressource by Duarte and Amaral~\cite{Duarte2018, Amaral2019}.

\section*{Acknowledgments}
This work is an extension of a PhD thesis chapter. It was supervised by P. Milman and A. Keller. We gratefully acknowledge helpful and fruitful discussions with S. Mansfield, A. Sainz, A. Cabello, M. Bitbol, L. de la Tremblaye, E. Dzhafarov, O. Gühne. G. M. acknowledges support from the French Agence Nationale de la Recherche (ANR-17-CE30-0006).

\clearpage
\bibliographystyle{unsrt}
\bibliography{contextualite}\vspace{0.75in}

\clearpage
\appendix
\onecolumngrid
\section{On the necessity to take into account noise on both states and observables}
\label{sub:dualitystatemeasurement}
In the modelisation of noise we considered in section~\ref{NoiseforPM}, noise has been applied only on states, not measurements. It is the usual way according to the equation $\moy{A} = \Tr \left( A \mathcal{E} (\rho)\right)$. We wondered whether it would ne becessary to yield also noise on the observables. 

\subsection{Equivalence between applying a quantum channel on a state or an observable if there is only one observable}
Applying one quantum channel on state is equivalent to applying its dual on observables if there is only one measurement. Indeed, with $\mathcal{E}:=\rho \rightarrow K_i \rho K_i^{\dagger}$ a completely positive quantum channel:
\begin{align}
    \Tr \left( A \mathcal{E} (\rho)\right) &= \Tr \left( A \sum_i K_i \rho K_i^{\dagger} \right) \\
    &= \sum_i \Tr \left( A  K_i \rho K_i^{\dagger} \right) \\
    &= \sum_i \Tr \left(K_i^{\dagger} A  K_i \rho  \right) \\
    &= \sum_i \Tr \left(G_i A  G_i^{\dagger} \rho  \right) \\
    &= \sum_i \Tr \left(\mathcal{E}^* (A)\rho  \right) 
    \label{eq:StateMeasDuality}
\end{align}
where $\forall i, G_i = K_i^{\dagger}$ and $\mathcal{E}^*:=\rho \rightarrow G_i \rho G_i^{\dagger}$. 
\par Now, we can wonder, what would bring the application of channels on both the state and the observable? Let us consider two quantum channels $\mathcal{E}_1$ and $\mathcal{E}_2$ whose associated Kraus operators are respectively $K_i$ and $G_j$. A simple recombination yields 

\begin{align}
    \Tr \left( \mathcal{E}_2(A) \mathcal{E}_1 (\rho)\right) &= \Tr \left(\sum_j G_j A G_j^{\dagger} \sum_i K_i \rho K_i^{\dagger} \right) \\
    &= \sum_i \sum_j \Tr \left( A H_{i,j} \rho H_{i,j}^{\dagger} \right) \\
    &= \Tr \left(\mathcal{E}_3 (\rho) A  \right) 
    \label{eq:DualNoise}
\end{align}
where $\mathcal{E}_3$ is a quantum channel defined by Kraus operators $H_{i,j} = G_j^{\dagger} K_i$. Hence, adding noise on the observable does not grant any more information. 

\subsection{No equivalence between applying quantum channels on states or observables in the multiple observables case?}

On a sequence of measurements, the noise models considered yield Eq~\eqref{eq:MeanAB}, that we reproduce for the comfort of the reader. The expectation value of the succession of observables A, B and C is 
\begin{align}
    \moy{ABC} &= \Tr \left( C \mathcal{E} \{ \Pi_B^+  \mathcal{E} ( \Pi_A^+ \rho \Pi_A^+ - \Pi_A^- \rho \Pi_A^-) \Pi_B^+  \right.\nonumber \\
    &- \left. \Pi_B^-  \mathcal{E} ( \Pi_A^+ \rho \Pi_A^+ - \Pi_A^- \rho \Pi_A^-) \Pi_B^- \} \right)
    \label{eq:MeanAB2C}
\end{align}
We are working on establishing whether it is still equivalent in this case to apply quantum channels on states or on observables.

%
%
%
%
%

\end{document}